\pdfoutput=1
\documentclass[11pt,twoside,a4paper,cmspaper,final,collab]{cms-tdr}
\def\svnVersion{c4dfea1}\def\svnDate{2023/09/17}\def\cmsCernNoTag{CERN-EP-2022-155}\def\cmsCernDate{\today}\def\cmsMessage{Published in Physics Letters B as \href{http://dx.doi.org/10.1016/j.physletb.2023.137703}{\doi{10.1016/j.physletb.2023.137703}.}}
\begin{document}\cmsNoteHeader{TOP-21-014}

\newcommand{\acc}{\ensuremath{\alpha\varepsilon}\xspace}
\newcommand{\AC}{\ensuremath{A_{\text{C}}}\xspace}
\newcommand{\deltay}{\ensuremath{\Delta\abs{y}}\xspace}
\newcommand{\ptmu}{\ensuremath{\pt^{\PGm}}\xspace}
\newcommand{\ptele}{\ensuremath{\pt^{\Pe}}\xspace}
\newcommand{\DRmin}{\ensuremath{\Delta R_{\text{min}}(\ell, j)}\xspace}
\newcommand{\ppbar}{\ensuremath{p\bar{p}}\xspace}

\newcommand{\ptrel}{\ensuremath{p_{\text{T},\text{rel}}}\xspace}
\newcommand{\pttop}{\ensuremath{p_{\text{T},\text{top}}}\xspace}

\newcommand{\ptrellj}{\ensuremath{\ptrel(\Pell,j)}\xspace}
\newcommand{\mttbar}{\ensuremath{M_{\ttbar}}\xspace}
\newcommand{\chisq}{\ensuremath{\chi^2}\xspace}

\newcommand{\mujets}{\ensuremath{\Pgm+\text{jets}}\xspace}
\newcommand{\ejets} {\ensuremath{\Pe+\text{jets}}\xspace}

\newcommand{\Wjets}{\ensuremath{\PW+\text{jets}}\xspace}
\newcommand{\Zjets}{\ensuremath{\PZ+\text{jets}}\xspace}
\newcommand{\gjets}{\ensuremath{\gamma+\text{jets}}\xspace}

\newcommand{\btag}    {\ensuremath{\PQb}\:\text{tag}    \xspace}
\newcommand{\btagged} {\ensuremath{\PQb}\:\text{tagged} \xspace}
\newcommand{\btagging}{\ensuremath{\PQb}\:\text{tagging}\xspace}
\newcommand{\thad}    {\ensuremath{\PQt_{\mathrm{h}}}   \xspace}
\newcommand{\tl}    {\ensuremath{\PQt_{\Pell}}   \xspace}
\newcommand{\ttag}    {\ensuremath{\PQt}\;\!\text{tag}    \xspace}
\newcommand{\Wtag}    {\ensuremath{\PW}\:\text{tag}    \xspace}

\newcommand{\Lchan}{\ensuremath{\mathcal{L}_k}\xspace}
\newcommand{\Nreco}{\ensuremath{N_{\text{reco}}}\xspace}
\newcommand{\Ngen}{\ensuremath{N_{\text{gen}}}\xspace}

\newcommand{\Aji}{\ensuremath{A_{ji}}\xspace}
\newcommand{\dunc}{\ensuremath{\delta_u}\xspace}
\newcommand{\prior}{\ensuremath{N(\vec\dunc)}\xspace}
\newcommand{\rpos}{\ensuremath{r_{\text{pos}}}\xspace}
\newcommand{\rneg}{\ensuremath{r_{\text{neg}}}\xspace}
\newcommand{\ACfid}{\ensuremath{\AC^{\text{fid}}}\xspace}

\newcommand{\figrefs}[1]{Fig.~\ref{#1}}
\newlength\cmsTabSkip\setlength{\cmsTabSkip}{1ex}
\ifthenelse{\boolean{cms@external}}{\providecommand{\cmsTable}[1]{#1}}{\providecommand{\cmsTable}[1]{\resizebox{\textwidth}{!}{#1}}}

\ifthenelse{\boolean{cms@external}}{\providecommand{\cmsLeft}{upper\xspace}}{\providecommand{\cmsLeft}{left\xspace}}
\ifthenelse{\boolean{cms@external}}{\providecommand{\cmsRight}{lower\xspace}}{\providecommand{\cmsRight}{right\xspace}}

\newcommand{\colspace}{\hphantom{3}}

\newcolumntype{R}{>{$}r<{$}}
\newcolumntype{L}{>{$}l<{$}}
\newcolumntype{M}{R@{$\;$}L}

\cmsNoteHeader{TOP-21-014} \title{Measurement of the
  \texorpdfstring{\ttbar}{ttbar} charge asymmetry in events with highly Lorentz-boosted
  top quarks in \texorpdfstring{$\Pp\Pp$}{pp} collisions  at  \texorpdfstring{$\sqrt{s} = 13\TeV$}{sqrt(s) = 13 TeV}}

\date{\today}

\abstract{ The measurement of the charge asymmetry in top quark pair
  events with highly Lorentz-boosted top quarks decaying to a single
  lepton and jets is presented. The analysis is performed using
  proton-proton collisions at $\sqrt{s}=13\TeV$ with the CMS detector
  at the LHC and corresponding to an integrated luminosity of
  138\fbinv.  The selection is optimized for top quarks produced with
  large Lorentz boosts, resulting in nonisolated leptons and
  overlapping jets. The top quark charge asymmetry is measured for
  events with a \ttbar invariant mass larger than 750\GeV and
  corrected for detector and acceptance effects using a binned maximum
  likelihood fit. The measured top quark charge asymmetry of
  $(0.42_{- 0.69}^{+ 0.64})\%$ is in good agreement with the
  standard model prediction at next-to-next-to-leading order in
  quantum chromodynamic perturbation theory with next-to-leading-order
  electroweak corrections. The result is also presented for two
  invariant mass ranges, 750--900 and ${>}900\GeV$.}

\hypersetup{%
pdfauthor={CMS Collaboration},%
pdftitle={Measurement of the ttbar charge asymmetry in events with highly Lorentz-boosted top quarks in pp collisions at sqrt(s) = 13 TeV},%
pdfsubject={CMS},%
pdfkeywords={CMS, top quark}
}

\maketitle 

\section{Introduction}
The vast majority of top quarks produced at hadron colliders are from
\ttbar pairs that originate via the strong interaction. At the LHC proton-proton ($\Pp\Pp$)
collider, the production stems from gluon fusion about 90\% of the time, with \qqbar
annihilation making up the rest~\cite{Czakon:2013goa,Catani:2019hip}.  
At leading order, the standard model
(SM) predicts that \ttbar production from \qqbar annihilation is
forward-backward symmetric. However, higher-order SM effects result in
a small (${\approx}6.6\%$) positive forward-backward asymmetry \AFB,
such that the top quark (antiquark) is preferentially emitted in the
direction of the incoming quark (antiquark). The \AFB was measured at the 
Tevatron proton-antiproton collider by the CDF and D0 collaborations~\cite{PhysRevLett.120.042001} and found to be 
consistent with the theoretical predictions~\cite{Czakon_2015}. There
is no asymmetry in the gluon fusion \ttbar production that dominates
at the LHC, but because valence quarks carry, on average, larger
momentum than antiquarks (from the sea), the rapidity distribution of
top quarks at the LHC is expected to be broader than that of top
antiquarks~\cite{Czakon_2018,Czakon_2016}. The \ttbar charge asymmetry
is defined as
\begin{linenomath*}
\begin{equation}
\AC = \frac{N(\deltay>0) - N(\deltay<0)}{N(\deltay>0) + N(\deltay<0)},
\label{eq:AC}
\end{equation}
\end{linenomath*}
where $\deltay = \abs{y_\PQt}-\abs{y_\PAQt}$ is the difference between the
absolute value of the top quark and antiquark rapidities and $N$ is
the number of events. The value of \AC is expected to be about 1\% in
the SM for LHC~\cite{Czakon_2018}.

Since the relative contribution of valence quarks increases at high
momentum transfer~\cite{PDF4LHC}, we expect that measuring \AC in a
sample of $\ttbar$ events with highly Lorentz-boosted top quarks will
lead to a more stringent probe of quantum chromodynamic (QCD)
predictions and higher sensitivity to beyond-the-SM (BSM) physics
processes that might alter the charge
asymmetry~\cite{arXiv:1109.3710}. Several models predict enhancements
with respect to the SM prediction in the presence of new particles,
including axigluons~\cite{arXiv:0709.1652, Frampton:2009rk}, \PZpr 
bosons~\cite{arXiv:9607207, arXiv:0809.3354, Ferrario:2009bz}, \PWpr
bosons~\cite{Aguilar-Saavedra:2011bsa,Ferrario:2009bz,Frampton:2009rk}, 
scalar isodoublets~\cite{Aguilar-Saavedra:2011mor}, 
color triplet scalars~\cite{Benbrik:2013lxa,Jung:2009pi}, and color sextet
scalars~\cite{Aguilar-Saavedra:2011bsa, Aguilar-Saavedra:2011mor}. These models
introduce new spin-0 and spin-1 particles in the interaction,
modifying \AC by exchanging the new particles through interference
terms and dedicated loops. 
The ATLAS and CMS
Collaborations have combined their inclusive and differential
measurements of \AC~\cite{ATLAS:2017gkv} at two center-of-mass
energies, obtaining $\AC = 0.005 \pm 0.007\stat \pm 0.006\syst$ and
$0.0055 \pm 0.0023\stat \pm 0.0025\syst$ at 7 and 8\TeV,
respectively.  These combined measurements show good agreement
with the respective SM predictions and uniquely restrict the phase
space of possible BSM phenomena that would produce large
asymmetries~\cite{Aguilar-Saavedra:2011bsa}.
Along with specific BSM models, deviations
from the SM prediction can also be interpreted through an effective
field theory (EFT) approach in which new physics contributions are
described via a fixed set of dimension-six operators added to the SM
Lagrangian~\cite{PhysRevD.83.034006}. In particular, fits to top quark production and differential distributions, together with Higgs boson,
boson pair (diboson), and electroweak precision measurements, are being used in global SM EFT analyses to 
constrain the operators~\cite{SMEFT,ATLAS:2022mlu}. 

This Letter presents the first measurement of the \ttbar charge
asymmetry that uses $\Pp\Pp$ collision data at $\sqrt{s} = 13\TeV$ and
optimizes the reconstruction of events with $\ttbar$ invariant mass
($\mttbar$) above 750\GeV. The kinematic requirements necessarily
imply highly boosted top quarks.  We target the single-lepton channel,
in which both top quarks decay as $\PQt \to \PQb \PW$, with one \PW
boson decaying leptonically ($\PW \to \Pell\PGn$, where $\Pell$ is
either a muon or electron), and the other hadronically ($\PW \to
\Pq\PAQq'$).  The charge of the lepton is used to distinguish the top quark from the 
top antiquark. The highly boosted top quarks yield collimated
decay products that appear as partially or fully overlapping energy deposits in our detector. 
For the top quark decaying to a \PW boson that decays leptonically (called a top quark
leptonic decay), the energy deposits from the lepton overlap with those of the 
\cPqb quark. Dedicated jet and lepton selection requirements at the trigger and 
offline levels~\cite{Sirunyan_2019}
allow us to reconstruct the decay products of the boosted leptonically
decaying top quarks by identifying the energy deposits of the leptons without requiring a 
minimum separation from other signals in the detector.
The overwhelming QCD multijet background is controlled with topological
requirements.  The topology of the top quark decaying to a \PW boson
that decays hadronically (called a top quark hadronic decay) depends
on the magnitude of its transverse momentum (\pt).  At the high end of
the \pt spectrum, the top quark decay products have angular distances
between partons that result in overlapping energy deposits. 
In contrast, at the low end of the \pt spectrum near the kinematic
threshold, the energy deposits from each parton appear separated from the others. For intermediate
\pt values, the energy deposits of partons originating from the hadronic \PW boson decay are overlapping, 
but the \PQb quark is identified
separately. All three topologies are considered in this analysis and
are referred to as ``merged'', ``semiresolved'', and ``resolved'' for
the high, intermediate, and low \pt regions, respectively.
Tabulated results are provided in the HEPData record for this analysis~\cite{hepdata}.

\section{The CMS detector and object reconstruction}

The central feature of the CMS apparatus is a superconducting solenoid
of 6\unit{m} internal diameter, providing a magnetic field of
3.8\unit{T}. Within the solenoid volume are a silicon pixel and strip
tracker, a lead tungstate crystal electromagnetic calorimeter, and a
brass and scintillator hadron calorimeter, each composed of a barrel
and two endcap sections. Forward calorimeters extend the
pseudorapidity ($\eta$) coverage provided by the barrel and endcap
detectors. Muons are measured in gas-ionization detectors embedded in
the steel flux-return yoke outside the solenoid. Events of interest
are selected using a two-tiered trigger
system~\cite{CMS:2020cmk,CMS:2016ngn}. A more detailed description of
the CMS detector, together with a definition of the coordinate system
used and the relevant kinematic variables, can be found in
Ref.~\cite{CMS:2008xjf}.

The offline event reconstruction is based on a particle-flow (PF)
algorithm~\cite{CMS:2017yfk}, which combines information from each
subdetector to identify muons, electrons, photons, and charged or neutral
hadrons. To avoid inefficiencies observed in the data for very high $\pt$ PF muons, we use muons determined by a different algorithm~\cite{CMS:2014pgm}, where muons are reconstructed first in the muon system and then fitted to tracks in the pixel and strip tracker. The primary
vertex is taken to be the vertex corresponding to the hardest
scattering in the event, evaluated using tracking information alone,
as described in Section 9.4.1 of Ref.~\cite{CMS-TDR-15-02}.  Charged
hadrons associated with other vertices are removed from further
consideration. The remaining PF candidates are clustered into jets
using the anti-\kt algorithm~\cite{Cacciari:2008gp,Cacciari:2011ma}
with a distance parameter of 0.4 (called AK4 jets). The same PF candidates are used to build
large-radius (AK8) jets using a distance parameter of 0.8, with the effects of additional 
$\Pp\Pp$ collisions in the same bunch crossing mitigated through the
pileup per particle identification algorithm (PUPPI)~\cite{Bertolini:2014bba,CMS:2020ebo}. 
Any AK4 jet with $\Delta R = \sqrt{\smash[b]{(\Delta\eta)^2+(\Delta\phi)^2}} <0.8$
from the closest AK8 jet, where $\phi$ is the azimuthal angle, is discarded from the event.  The
total jet \ptvec is given by the sum of the \ptvec of its
constituents.  If a lepton is found within $\Delta R < 0.4$ of an AK4
jet or ${<}0.8$ of an AK8 jet, its four-momentum is subtracted from
that jet~\cite{Sirunyan_2019}.  The missing transverse momentum vector
\ptvecmiss is computed as the negative vector sum of the transverse
momenta of all the PF candidates in an event, and its magnitude is
denoted as \ptmiss~\cite{CMS:2019ctu}.  Corrections are applied to
improve the jet energy scale and resolution, and the \ptvecmiss is
modified to account for these corrections~\cite{CMS:2011shu,Khachatryan_2017}.

Specialized techniques use AK8 jets and jet substructure
information~\cite{CMS:2020poo}, including ``soft-drop
clustering''~\cite{Larkoski:2014wba} and
``N-subjettiness''~\cite{Thaler:2010tr}, to identify the hadronic decay of 
boosted top quarks, following the techniques detailed in Ref.~\cite{Sirunyan_2019}. 
Two exclusive categories are considered: 
hadronically decaying top quarks ($\ttag$) in which the three partons
are merged into a single AK8 jet, and hadronically decaying \PW bosons
($\Wtag$) in which the two partons from the \PW boson are merged into
a single AK8 jet, but the bottom quark is reconstructed as a separate
AK4 jet. The identification of jets originating from the decay of \PQb
hadrons ($\btag$) employs a deep neural network multi-classification
algorithm (\textsc{DeepJet}~\cite{CMS:2017wtu}) that relies on
information from the tracker and the
calorimeters~\cite{DeepJetJINST2020}. The $\btagging$ algorithm is
applied to each AK4 jet $j$ with $\pt^{j}>50\GeV$ and
$\abs{\eta^{j}}<2.4$. The \PQt and \PW tagging algorithms are applied to
AK8 jets with $\pt^{j}>400\GeV$ and $\abs{\eta^{j}}<2.4$. The algorithms
are tuned for misidentification rates smaller than $5\%$ and have efficiencies
greater than 80\%.

\section{Collider data and simulated samples}

We analyze data collected by the CMS detector during Run 2
(2016--2018) and corresponding to a total integrated luminosity of 
138\fbinv~\cite{CMS-LUM-17-003, CMS-PAS-LUM-17-004,
  CMS-PAS-LUM-18-002}.  Events in the muon channel ($\mujets$) are
selected with a single-muon trigger that requires $\ptmu>50\GeV$.
Events in the electron channel ($\ejets$) are selected by either a
single-electron trigger with $\ptele>115\GeV$ or a trigger requiring
one electron with $\ptele>50\GeV$ and one jet with
$\pt^j>165\GeV$~\cite{Sirunyan_2019}. The efficiency of these triggers for events in our signal 
sample is above $95\%$. As the $\ejets$ trigger was not
available during the early running period in 2017, the integrated
luminosity available for the 2017 $\ejets$ channel is reduced by
5\fbinv.

In the offline reconstruction, we select events for the $\mujets$
($\ejets$) channel that contain exactly one muon with $\ptmu>55\GeV$
and $\abs{\eta^\PGm}<2.4$ (one electron with $\ptele>80\GeV$ and
$\abs{\eta^\Pe}<2.5$) and at least two AK4 jets, $j_1$ and $j_2$, with
$\pt^{j_1}>150$ (185)\GeV, $\pt^{j_2}>50\GeV$, and
$\abs{\eta^{j_{1,2}}}<2.4$. To preserve the identification efficiency of
$\ttbar$ decay products in the highly boosted topology, no isolation
requirement is imposed on the leptons at either the trigger or offline
level. To reduce the background from QCD multijet events, we apply a
two-dimensional (2D) selection that requires leptons to satisfy either
the condition $\DRmin > 0.4$ or $\ptrellj > 25\GeV$, where \DRmin is
the angular separation between the lepton and the closest AK4 jet, and
\ptrellj is the transverse momentum of the lepton with respect to the
axis of the nearest AK4 jet~\cite{Sirunyan_2019}.  Finally, events
need to satisfy $\ptmiss>50\GeV$ and $\ptmiss + \ptmu>150\GeV$
($\ptmiss>120\GeV$) in the $\mujets$ ($\ejets$) channel. The larger
value of the $\ejets$ $\ptmiss$ requirement efficiently reduces the
larger QCD multijet background in this channel and precludes the need
for a separate requirement on $\ptmiss + \ptele$. To suppress the
contribution from the $\Wjets$ background, at least one of the AK4
jets must be $\btagged$.

A variety of Monte Carlo (MC) simulated samples are used to model the signal 
and background contributions. A comprehensive description of the generation 
parameters used, as well as the MC normalization and associated uncertainties, 
can be found in Ref.~\cite{CMS:2021vhb}. The $\ttbar$ and
single top quark (ST) processes are produced with the
next-to-leading-order (NLO) \POWHEG~\cite{Frixione:2007nw,POWHEG} generator. Simulated
\Wjets, Drell--Yan (DY) $\PZ/\gamma+\text{jets}$, and QCD multijet
processes are generated with \MGvATNLO~\cite{Alwall:2014hca} at
NLO. The contribution from diboson events is negligible. 
All samples are interfaced to \PYTHIA{8}~\cite{Sjostrand:2014zea} 
with the CP5 tune~\cite{Sirunyan_2020}
for parton showering and hadronization, and processed 
through a \GEANTfour-based simulation~\cite{Geant4}, which models the propagation 
of the particles through the CMS detector and the corresponding detector response.
The NNPDF 3.1 parton distribution functions
(PDFs)~\cite{NNPDF:2014otw} are used for all samples, and events in all samples include the
simulation of additional inelastic $\Pp\Pp$ interactions (pileup)
within the same or nearby bunch crossings. Small corrections are
applied to all MC samples to improve the agreement with the observed
data, derived from data control samples that are independent of the
candidate selection. In particular, the transverse momentum of the top
quarks at the generator level is multiplied by the function $e^{(0.0615-0.0005\pttop)}$ to match the distribution
measured in data~\cite{CMS:2021vhb}.

\section{Reconstruction of the top quark pair events}\label{sec:reco}

The $\ttbar$ system is reconstructed by assigning the four-vectors of
the final-state objects to either the leptonic ($\tl$) or hadronic
($\thad$) leg of the $\ttbar$ decay. The events are separated into the
three topologies discussed earlier based on the presence of \PQt- or \PW-tagged jets: 
the merged topology 
contains events with one $\ttag$ and no $\Wtag$; the semiresolved topology contains
events with one $\Wtag$ and no $\ttag$; and the resolved topology contains the rest
of the events that have no \PQt and no \PW tag.  Events with more than one 
\PQt or \PW tag are discarded. For events with a
$\ttag$, the \PQt-tagged jet is taken as the $\thad$ and only AK4 jets
with $\DR > 0.8$ from the $\thad$ are considered as candidates for the
$\tl$. For events with a $\Wtag$, the \PW-tagged jet is assigned to the
$\thad$. The AK4 jets with $\DR > 0.8$ from the $\Wtag$ can be
assigned to either the $\tl$ or the $\thad$, and at least one AK4 jet on each side is required. 
For events with neither a $\ttag$ nor a $\Wtag$, all possible assignments of AK4 jets are
considered for both the $\tl$ and the $\thad$, and again, at least one AK4 jet on each side is required. 
There is no upper limit on the number of AK4 jets in any of the topologies.  Even though each event
includes at least one \PQb-tagged jet, the $\btagging$ information for
individual jets is not used. The longitudinal
component of the neutrino momentum is inferred by constraining the
invariant mass of the lepton plus neutrino system to the \PW boson
mass~\cite{JHEP07}. Assuming the \PW boson is on-shell, a quadratic
equation for the $z$ component of the momentum of neutrino can be derived:
 \begin{linenomath*}
\begin{equation}
 p_{z,\nu}^{\pm} = \frac{\mu p_{z,\ell}}{p_{\mathrm{T},\ell}^{2}} \pm
                   \sqrt{  \frac{\mu^{2} p_{z,\ell}^{2}}{p_{\mathrm{T},\ell}^{4}} - \frac{E_{\ell}^{2}p_{\mathrm{T},\nu}^{2}
                         - \mu^{2}}{p_{\mathrm{T},\ell}^{2}}}~,
 \label{eq:Pz2}
\end{equation}
 \end{linenomath*}
where $p_{\ell}$ and $p_{\nu}$ are the three momenta of the charged lepton and the neutrino, respectively, 
with $\vec{p}_{T,\nu} = \ptvecmiss$, and 
$\mu = \frac{1}{2}M_{W}^{2} + \vec{p_{\mathrm{T},\ell}} \cdot \vec{p_{\mathrm{T},\nu}}$.
Equation~(\ref{eq:Pz2}) has either 0, 1, or 2 real solutions.
In the absence of a real solution,
the real part of the complex solutions is used.
If there are two real solutions,
both cases are tested,
effectively doubling the number
of hypotheses for that event.

\begin{figure*}[bt]
\centering
\includegraphics[width=0.49\textwidth]{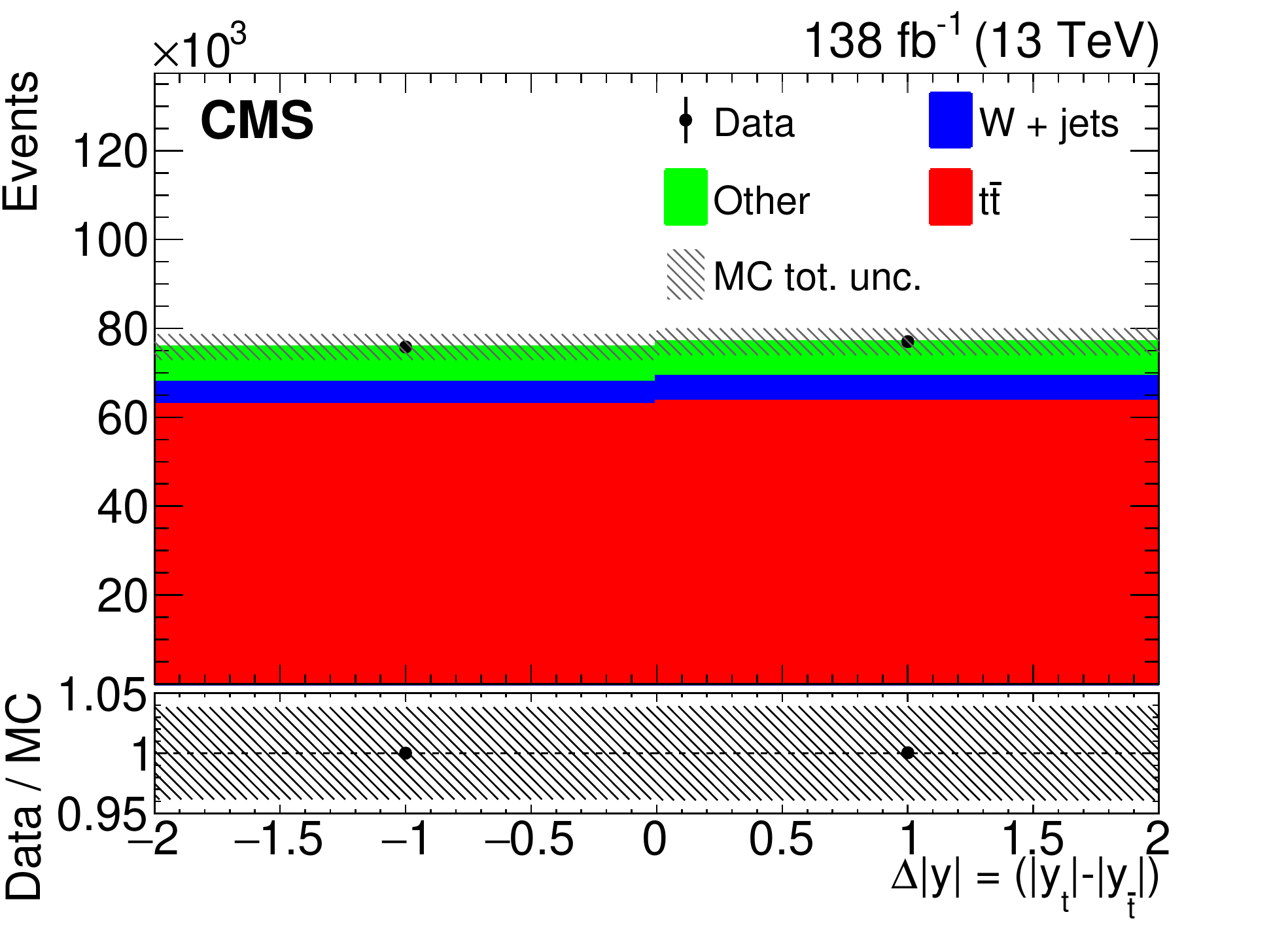}
\includegraphics[width=0.49\textwidth]{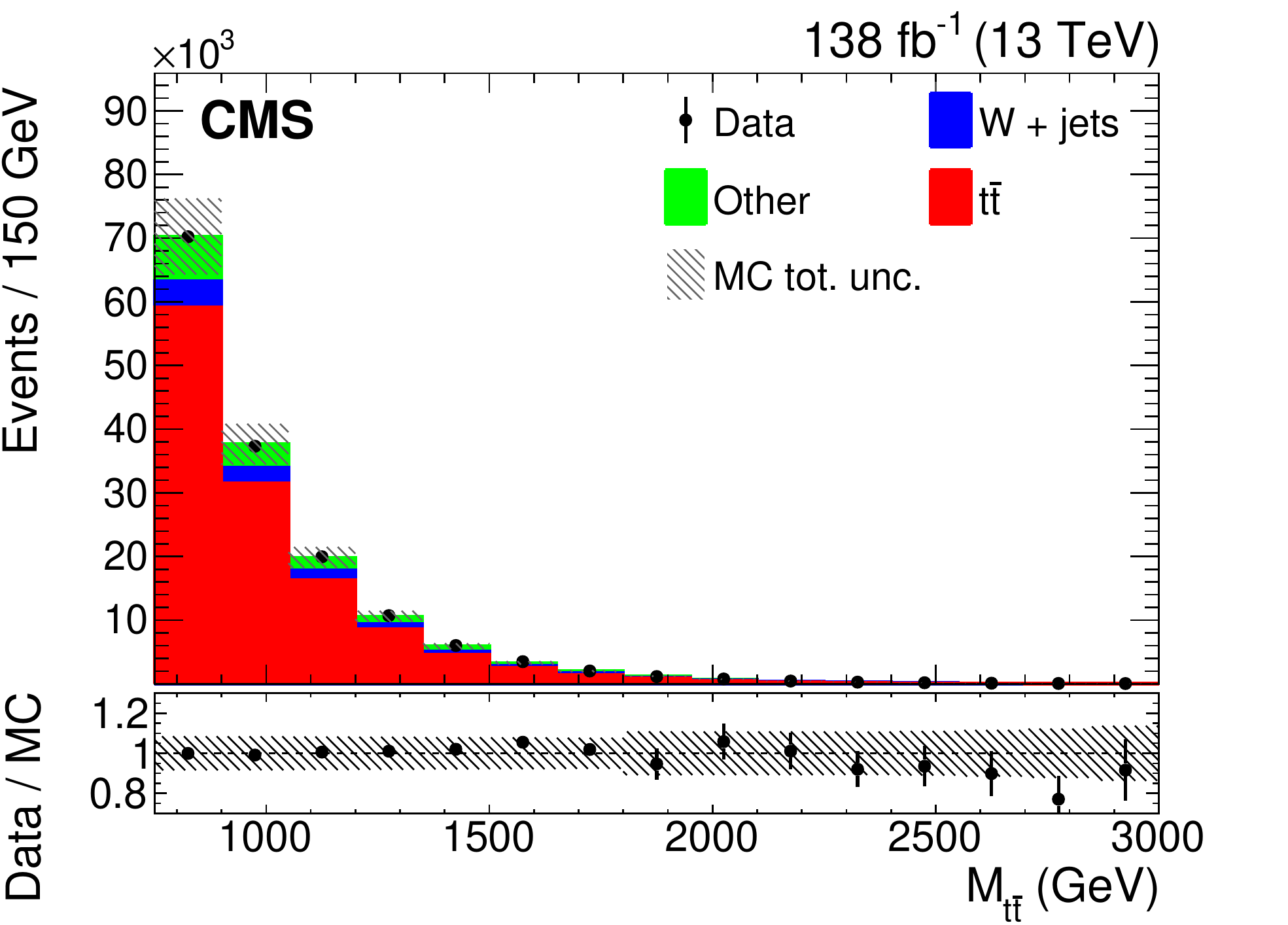}
\includegraphics[width=0.49\textwidth]{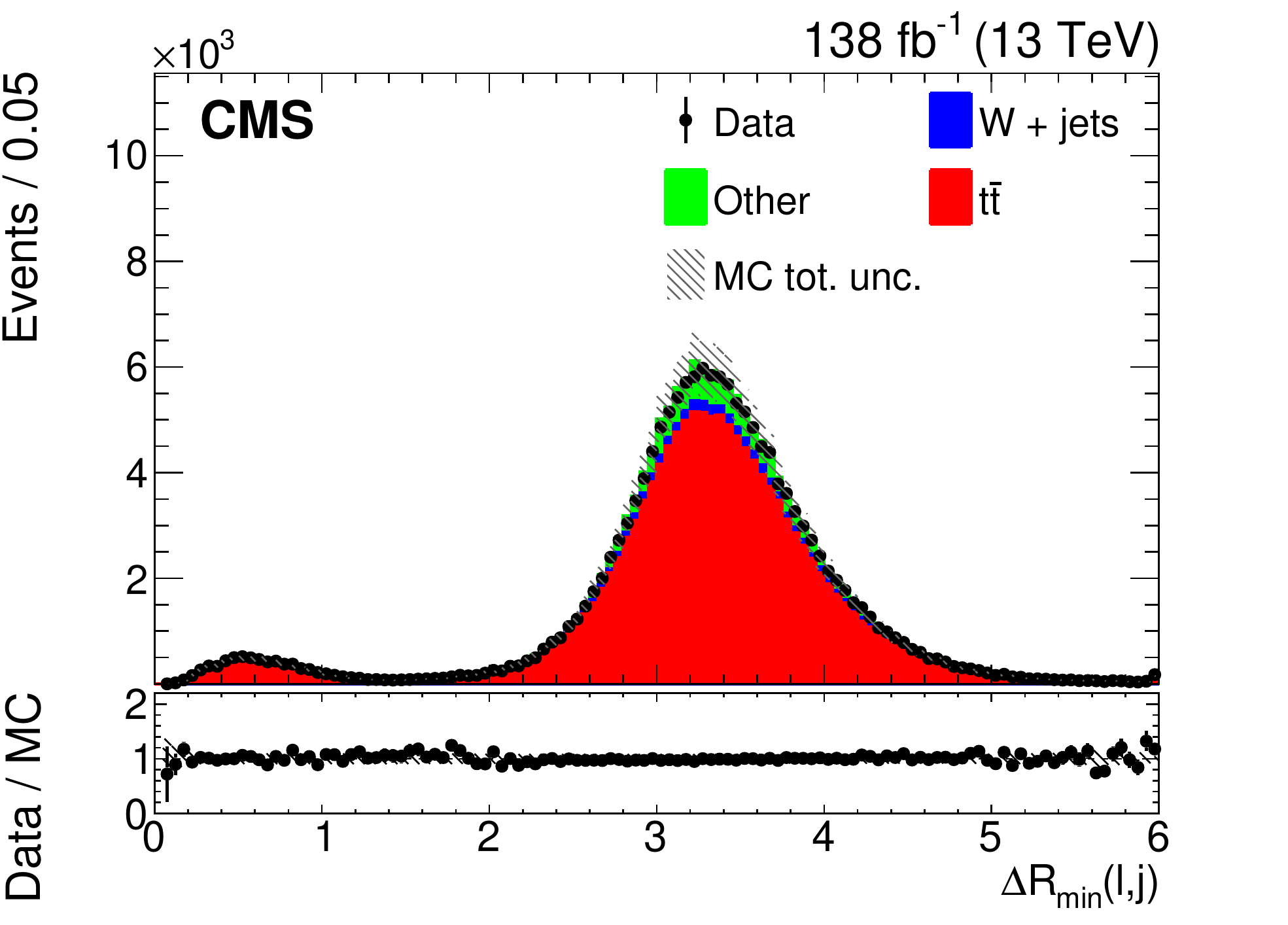}
\includegraphics[width=0.49\textwidth]{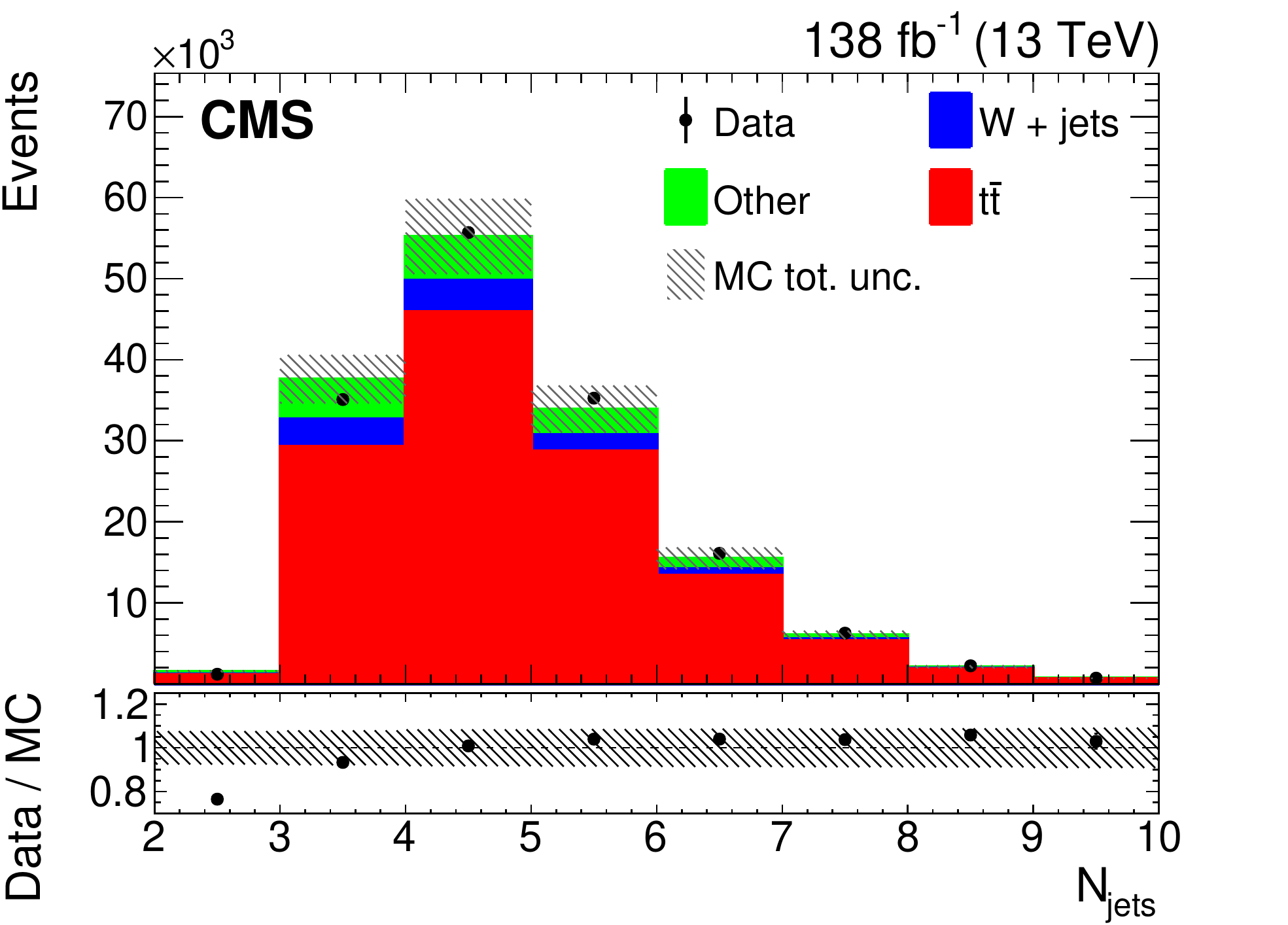}
\caption{Comparison between data and MC simulation for kinematic
  distributions based on events in the candidate sample
  (described in Section~\ref{sec:reco}): \deltay (upper left),
  reconstructed $\mttbar$ (upper right), distance between the lepton
  and the closest AK4 jet $\DRmin$ (lower left), and the number of
  AK4 jets (lower right). The vertical bars on the points show the
  statistical uncertainty in the data. The shaded bands represent the
  total uncertainty in the MC predictions (described in
  Section~\ref{sec:syst}). The lower panels give the ratio of the data
  to the sum of the MC predictions. The distributions are shown 
  after the likelihood normalization described in Section~\ref{sec:unfold}. 
\label{fig:Kinematics}}
\end{figure*}

For each event, one $\ttbar$ hypothesis is selected based on a two-term 
 \chisq discriminator used to quantify the compatibility of each hypothesis with 
 a \ttbar decay. The discriminator, which was optimized for \ttbar invariant masses greater than 750\GeV, 
 is defined as 
 \begin{linenomath*}
\begin{equation}
 \chi^{2} =   \left [ \frac{M_\text{lep} - \langle{M}_\text{lep}\rangle}{\sigma_{M_\text{lep}}} \right ]^2
            + \left [ \frac{M_\text{had} - \langle{M}_\text{had}\rangle}{\sigma_{M_\text{had}}} \right ]^2,
 \label{eq:chi2}
\end{equation}
 \end{linenomath*}
where $M_{\text{lep}}$  and $M_\text{had}$ are the invariant masses
of the reconstructed $\tl$ and $\thad$ decaying top quark, respectively. The quantities
$\sigma_{M_{\text{lep}}}$ and  $\sigma_{M_\text{had}}$ are the resolutions of the leptonic and hadronic top quark 
reconstruction, respectively, and $ \langle{M}_\text{lep}\rangle$ and $\langle{M}_\text{had}\rangle$ are the means of the 
corresponding mass distributions obtained from simulation for each of the topologies.
Because background processes typically
result in large values of \chisq, events with \chisq$>30$ are rejected. 
Finally, only events with $\mttbar > 750\GeV$ are retained in
our signal candidate sample.   
Figure~\ref{fig:Kinematics} shows comparisons between data and MC
simulation for kinematic distributions based on events in the
candidate sample. The distributions are shown 
after the likelihood normalization described in Section~\ref{sec:unfold}. 
The boosted nature of the top quarks in the events
becomes evident: the $\mttbar$ range extends to multi-\TeVns values,
events with two and three AK4 jets originating from the collimated top
quark decay products are reconstructed, and events with leptons next to the jet axis are retained.
Generally good agreement
between data and MC is observed.

\section{Systematic uncertainties}\label{sec:syst}

Systematic uncertainties from numerous sources can affect the
normalization and the shape of the distributions of physical
observables in both signal and background samples.  The systematic
uncertainties affecting only the normalization come from the SM
theoretical cross section values for each process and the luminosity
normalization. All MC samples are normalized according to their
respective SM cross section values, as has been done in previous analyses including 
Ref.~\cite{CMS:2021vhb}, and assigned a rate uncertainty of
30\% for background processes~\cite{CMS:2019prd}  and 5\% for the \ttbar
signal~\cite{CZAKON20142930}. Additionally, uncertainties in the
integrated luminosity vary from year to year: 2.5, 2.3, and 1.2\% for
2018~\cite{CMS-PAS-LUM-18-002}, 2017~\cite{CMS-PAS-LUM-17-004}, and
2016~\cite{CMS-LUM-17-003}, respectively. These include both correlated
and uncorrelated components across the three years, while the overall uncertainty for the 2016--2018 period is 1.6\%.

All other systematic uncertainties affect both the normalization and
shape of the MC distributions.  Uncertainties from experimental
sources are applied to both signal and background samples.  All MC
samples are reweighted to match the pileup distribution in data, which
is generated by using the instantaneous luminosity per bunch crossing
for each luminosity section, with a total inelastic cross section of
69.2\unit{mb}; an uncertainty of 4.6\% is applied to this
value~\cite{CMS:2018mlc}. All muons and electrons in the simulated
samples have uncertainties associated with the high-level trigger
(HLT), reconstruction (reco), and identification (ID). The uncertainty associated with the possible misidentification
of the sign of the lepton electric charge is negligible. These
uncertainties are uncorrelated across lepton flavors but correlated
across years and are parameterized as a function of the \pt and $\eta$
of the leptons.  There is a uniform uncertainty in the efficiency of
the 2D selection that rejects QCD background, and this is uncorrelated
across lepton flavors and years.  Uncertainties in the jet energy
corrections (JEC) and resolution (JER) are parameterized in terms of
the jet \pt and $\eta$ and considered correlated across years.  The
uncertainties in the tagging scale factors are parameterized as a
function of the jet \pt. The uncertainties in \PQt and \PW tagging are
100\% correlated across years, but the uncertainty in \btagging has
both correlated and uncorrelated components~\cite{CMS-DP-2018-058}.
There are different scale factors to account for the cases when the
tagging algorithms incorrectly identify (mistag) some jets, so a
separate mistagging uncertainty is also assigned.

In addition to the experimental sources, we consider uncertainties
affecting the MC simulations. The uncertainty from the choice of PDF is estimated from the Hessian NNPDF3.1 sets 
according to the procedure described in Ref.~\cite{NNPDF:2017mvq}.  
Renormalization ($\mu_R$) and factorization ($\mu_F$) scales at the matrix element level are varied
by a factor of 2 or 0.5 to take into account the effect of
higher-order corrections in the \ttbar and \Wjets simulations. The
matrix element and parton shower matching scale ($h_{\text{damp}}$)
regulates the high-\pt radiation by damping real emissions from the
\POWHEG generator; this effect is only taken into account for \ttbar
and evaluated using independent simulated samples generated based 
on parameters derived in Ref.~\cite{Sirunyan_2020}. Uncertainties
related to the modeling of initial- and final-state radiation (ISR and
FSR) are determined by varying the strong
coupling constant \alpS at the scale $Q^2$ for the \ttbar
samples.  Finally, an uncertainty in the correction to the top quark
\pt is evaluated as a one-sided variation computed from the difference
between the top quark $\pt$ distribution with and without the
correction~\cite{CMS:2021vhb}. For all these uncertainties, those
originating from the same source are considered as 100\% correlated
between channels and those arising from different sources are
considered to be 100\% uncorrelated.

\section{Unfolded results}\label{sec:unfold}

The top quark charge asymmetry is obtained by performing a
simultaneous binned maximum likelihood fit to data in all bins and
categories of the candidate sample as implemented by the LHC Higgs Combination Group~\cite{Combine}.
Statistical uncertainties
due to the MC sample size are treated separately in each bin
with the Barlow--Beeston-lite approach~\cite{BARLOW1993219}. Each
source of systematic uncertainty is included in the likelihood as a
unique nuisance parameter.  For contributions that apply to multiple
analysis channels, the nuisance parameters are fully correlated,
allowing better constraints to be placed on the systematic
uncertainties. For a given channel $k$ in our
analysis, the corresponding likelihood function \Lchan is defined as: 
\begin{linenomath*}
\begin{equation}
\label{eq:likelihood}
\Lchan = \prod_{j=1}^{\Nreco} P\left( n_j ; \sum_{i=1}^{\Ngen} \Aji(\vec{\delta_u})\mu_{i}(\vec{\delta_u}) + b_j(\vec{\delta_u}) \right) \prior,
\end{equation} 
\end{linenomath*}
where
\begin{itemize}
\item $P\left( n ; \mu \right)$ represents the Poisson probability of observing $n$ events when $\mu$ are expected.

\item The indexes $i$ and $j$ run over the number of bins at generator level (\Ngen) and reconstruction level (\Nreco), respectively.  In this
  analysis, we use two bins $(\Nreco=\Ngen=2)$ corresponding to the positive (bin 1) and negative (bin 2)
  difference between the absolute value of the top quark and antiquark
  rapidities \deltay.
  
\item $\vec{\delta_u}$ are the nuisance parameters.
 
\item \Aji is the response matrix, which gives the probability for an event
reconstructed in bin $j$ to have been produced in bin $i$.  It is implemented by including the relevant number of
  reconstructed and generated simulated \ttbar events for each entry, which are subject to the effects of the
  nuisance parameters. This implementation allows the matrix to account for effects from detector resolution (smearing) as well as detector acceptance and efficiency.

\item  $\mu_{1}=\rpos N^\text{gen}_\text{pos}$ and $\mu_{2}=\rneg N^\text{gen}_\text{neg}$, where \rpos and \rneg are the signal strengths multiplying
  the number of signal events at generator level with $\mttbar^\text{gen} > 750\GeV$ in which the value of $\deltay^\text{gen}$ is positive ($N^\text{gen}_\text{pos}$) or negative ($N^\text{gen}_\text{neg}$), respectively.

\item $n_j$ is the number of data events in bin $j$.

\item $b_j$ is the number of background events predicted in bin $j$.

\item \prior are the priors for the nuisance parameters, with the normalization and shape uncertainties assigned a log-normal and Gaussian distribution, respectively. The shape uncertainty nuisance parameters control the vertical interpolation of histograms that represent the up and down one standard deviation shifts from the nominal distribution ~\cite{Combine-github}. 
\end{itemize}

Each analysis channel is defined based on reconstruction-level quantities by a range of \mttbar values and a specific
year and lepton flavor. To account for migration of events between mass bins, all simulated events are included, even if
the generated mass is outside the reconstructed mass range under consideration.  The total likelihood is given by the
product of the individual likelihoods from Eq.~(\ref{eq:likelihood}), with the index $k$ running over all 12 channels:
two lepton flavors ($\mujets$ and $\ejets$), 3 years (2018, 2017, and 2016), and two mass regions ($750 < \mttbar <
900\GeV$ and $\mttbar > 900\GeV$).  This unfolding approach also has the advantage that the background contributions are
constrained by the fit, resulting in smaller systematic uncertainties than those obtained with a direct background
subtraction.  

The unfolded charge asymmetry at parton level is given by
 \begin{linenomath*} 
\begin{equation}
  A_C = \frac{\rpos N^\text{gen}_\text{pos} - \rneg N^\text{gen}_\text{neg}}{\rpos N^\text{gen}_\text{pos}+ \rneg N^\text{gen}_\text{neg}},
  \label{eq:ACfull}
  \end{equation}
   \end{linenomath*}
with $r_\text{pos}$ and $r_\text{neg}$ as free parameters in the fit.  However, to ensure the uncertainties on $A_C$ are properly estimated, we define $r_\text{pos}$ in terms of $r_\text{neg}$ and $A_C$, and select $r_\text{neg}$ and $A_C$ as the free parameters. The $A_C$ can also be measured in the two mass regions separately. This is achieved combining subsets of the 12 analysis channels according to the reconstructed \mttbar, while 
redefining $N^\text{gen}_\text{pos}$ and $N^\text{gen}_\text{neg}$ as the corresponding number of generated signal events with $750 < \mttbar^\text{gen} <900\GeV$ or $\mttbar^\text{gen} > 900\GeV$.  

\begin{figure*}[tbhp]
\centering
\includegraphics[width=0.49\textwidth]{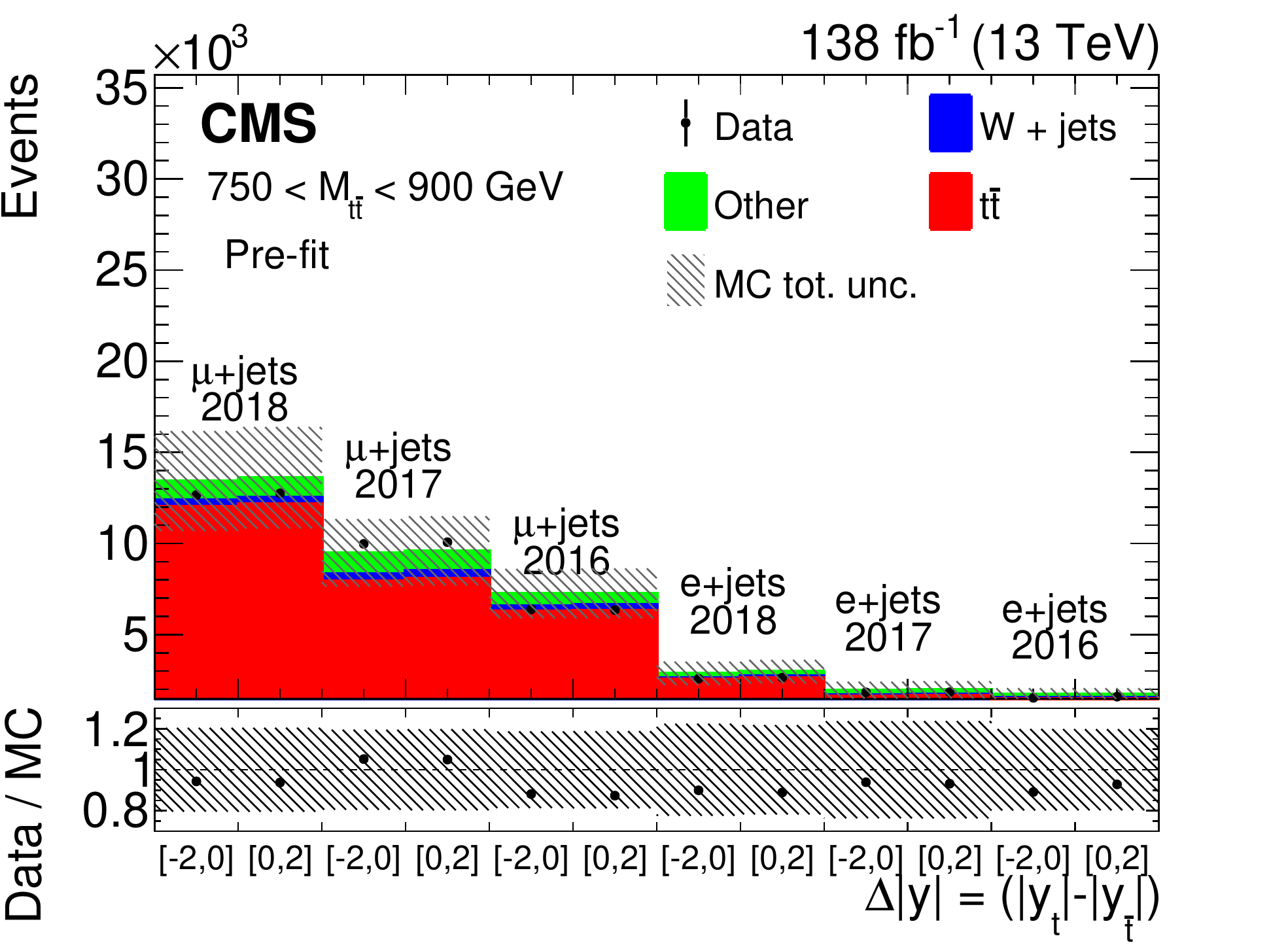}
\includegraphics[width=0.49\textwidth]{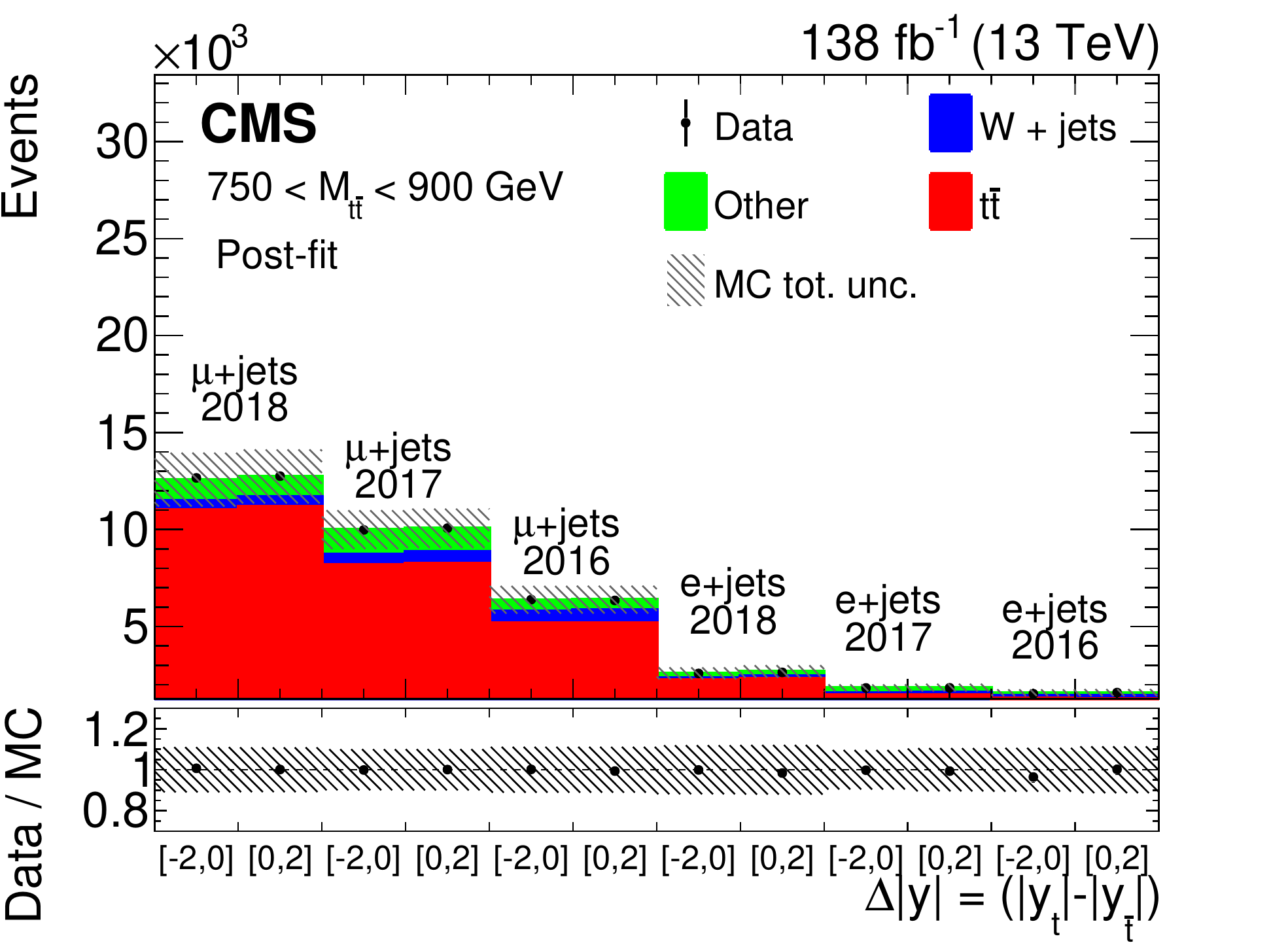}
\includegraphics[width=0.49\textwidth]{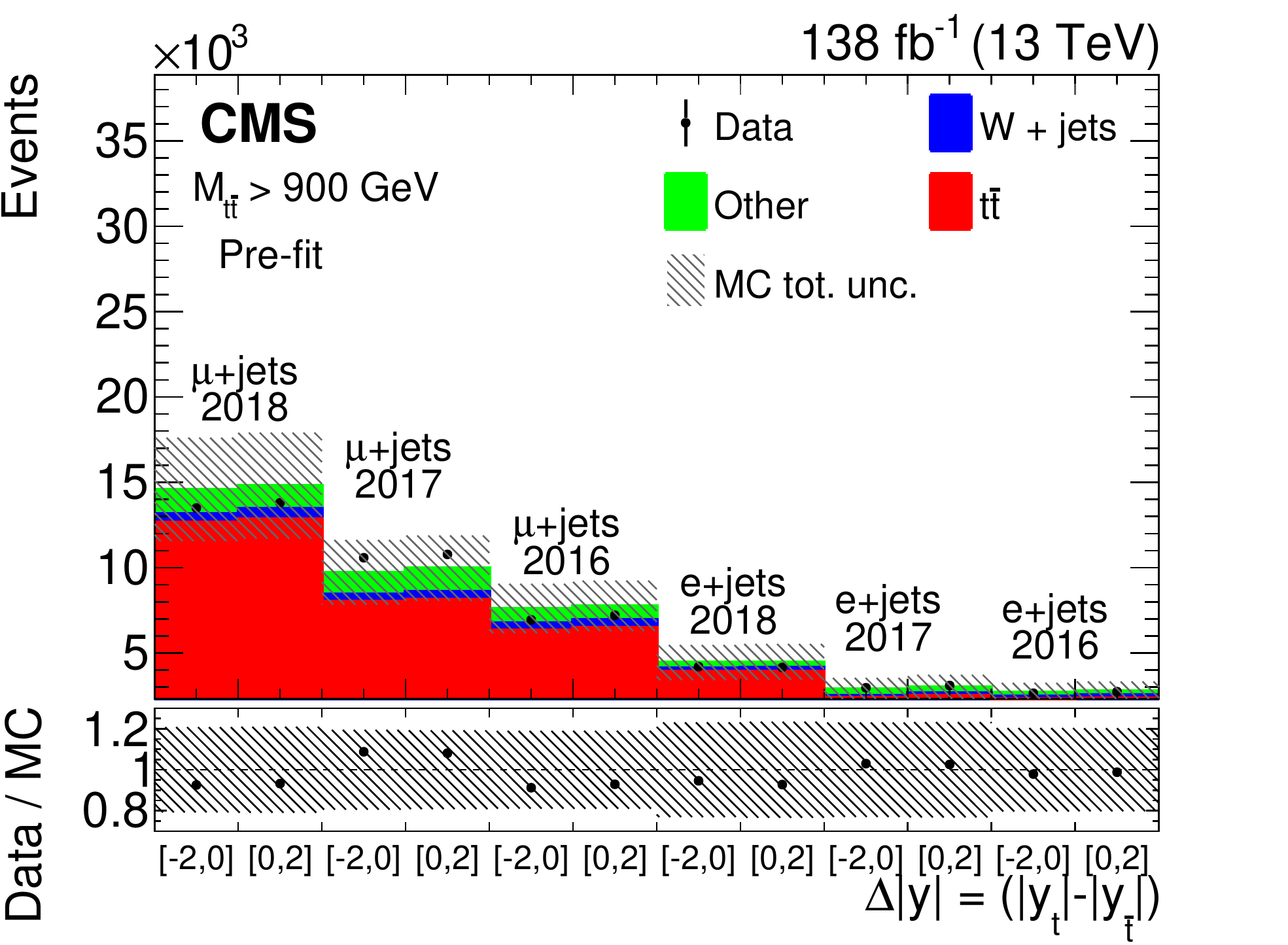}
\includegraphics[width=0.49\textwidth]{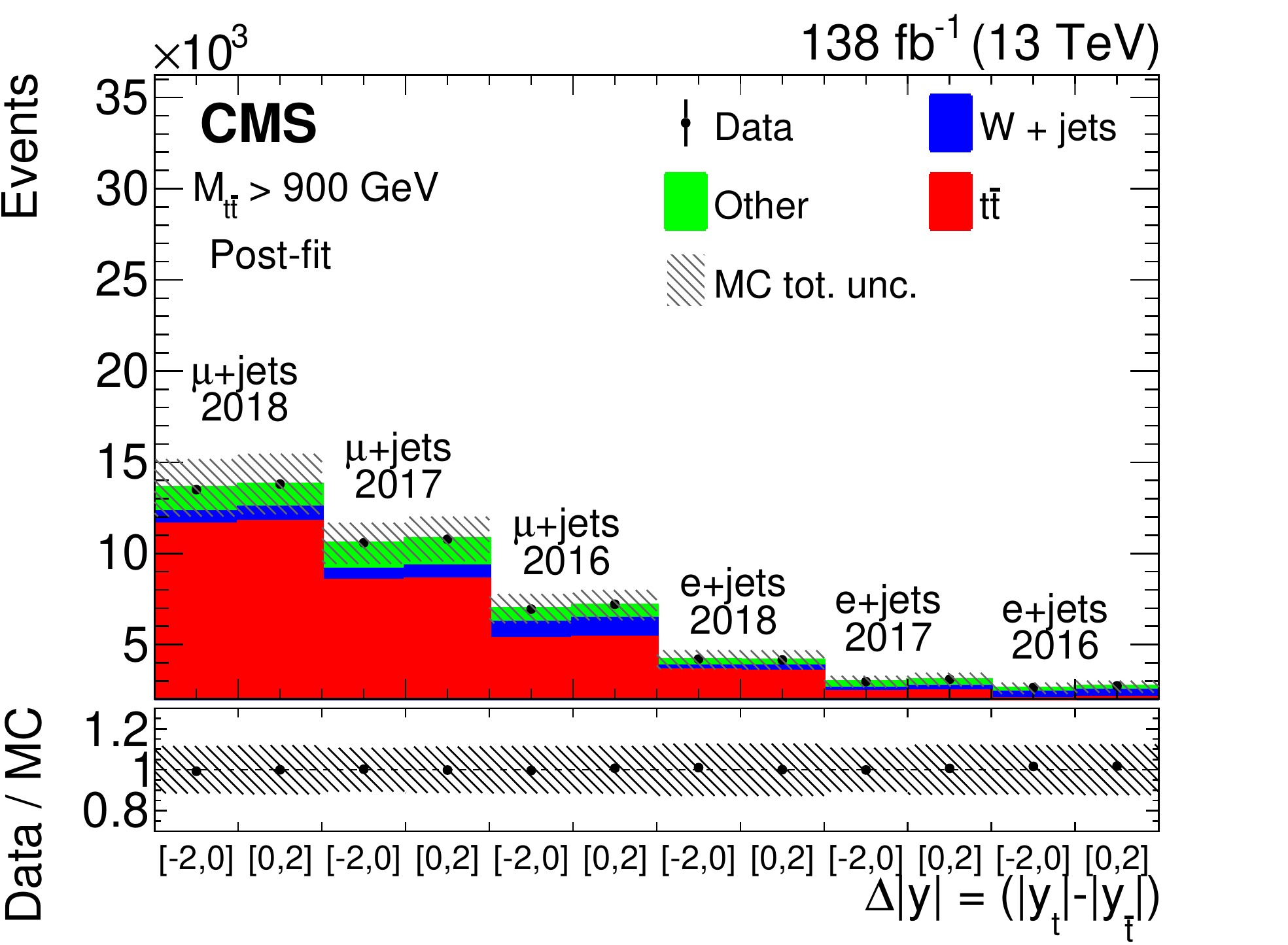}
\caption{Comparison between data and MC simulation for $\deltay$ for
  each of the 12 analysis channels, both before (left) and after
  (right) the likelihood normalization. The plots in the upper row
  correspond to $750 < \mttbar < 900\GeV$, and the plots in the lower
  row to $\mttbar > 900\GeV$. The vertical bars on the points
  represent the statistical uncertainties in the data and the shaded
  bands give the combined MC statistical and systematic
  uncertainties. The lower panels display the ratio of the data yields
  to the sum of the MC predictions.
\label{fig:Deltay}}
\end{figure*}

Table~\ref{tab:yields} shows the signal and background yields in our
candidate sample after the likelihood fit, separated into the
two mass regions.  The contributions to our candidate sample from
background processes (ST, $\Wjets$, DY, and QCD multijet) are taken
from simulation and their normalization allowed to change during the
likelihood fit. The higher $\pt$ thresholds on the
lepton and leading jet in the electron channel result in
significantly reduced signal acceptance compared to the muon channel.  
Figure~\ref{fig:Deltay} shows \deltay for
each of these 12 channels both before and after the likelihood
normalization. As can be observed, the likelihood fit reduces the
total uncertainty significantly and improves the agreement between
data and the MC prediction.

\newcolumntype{P}{@{$\,($}r@{$)$}}
\newcolumntype{S}{r@{${}\pm{}$}r}
\newcolumntype{H}{r@{\hphantom{${}\pm{}$}}}
\begin{table*}[tbh]
\centering
\topcaption{The event yields for the candidate sample in data and MC simulations after
  the likelihood fit for each of the 12 channels ($\mujets$, $\ejets$, 
  3 years; and two mass regions).  The uncertainties in the
  MC predictions include both statistical and systematic
  components.}
\renewcommand{\arraystretch}{1.1}
\cmsTable{
\begin{tabular}{lSSSSSS}
Process &  \multicolumn{2}{c}{$\PGm~(2018)$} & \multicolumn{2}{c}{$\PGm~(2017)$} & \multicolumn{2}{c}{$\PGm~(2016)$} & \multicolumn{2}{c}{$\Pe~(2018)$} & \multicolumn{2}{c}{$\Pe~(2017)$} & \multicolumn{2}{c}{$\Pe~(2016)$} \\
\hline
\multicolumn{13}{c}{} \\
\multicolumn{13}{c}{$750 < \mttbar < 900\GeV$}\\ 
\hline
$\ttbar$ & 22\,230 & 1950  & 16\,430 & 1400 & 10\,370 & 970 & 4590 & 450 & 2950 & 260 & 2560 & 240   \\ 
ST & 1620 & 150  & 2150 & 190 & 910 & 80 & 410 & 30 & 510 & 40 & 290 & 30   \\ 
$\Wjets$ & 970 & 110  & 1150 & 120 & 1250 & 270 & 240 & 30 & 220 & 30 & 320 & 70   \\ 
DY & 90 & 20  & 40 & 10 & 50 & 10 & 15 & 3 & 6 & 1 & 10 & 2   \\ 
QCD multijet & 410 & 100  & 270 & 60 & 180 & 40 & 6 & 2 & 2 & 1 & 0 & 0   \\ 
\hline
Total & 25\,310 & 1960  & 20\,050 & 1400 & 12\,770 & 1000 & 5270 & 450 & 3690 & 260 & 3180 & 250   \\ 
Data & \multicolumn{1}{H}{25\,417} & & \multicolumn{1}{H}{20\,052} & & \multicolumn{1}{H}{12\,735} & & \multicolumn{1}{H}{5219} & & \multicolumn{1}{H}{3674} & & \multicolumn{1}{H}{3127} & \\[\cmsTabSkip]
 \multicolumn{13}{c}{$\mttbar > 900\GeV$}\\ 
 \hline  
$\ttbar$ & 23\,340 & 2270  & 17\,120 & 1640 & 10\,700 & 1060 & 7140 & 740 & 4880 & 490 & 4110 & 430   \\ 
ST & 1650 & 140  & 2020 & 170 & 920 & 80 & 610 & 60 & 690 & 60 & 420 & 40   \\ 
$\Wjets$ & 1450 & 170  & 1330 & 160 & 1970 & 450 & 520 & 60 & 440 & 50 & 740 & 160   \\ 
DY & 110 & 20  & 60 & 10 & 70 & 10 & 30 & 6 & 14 & 3 & 20 & 4   \\ 
QCD multijet & 860 & 120  & 810 & 130 & 470 & 70 & 10 & 3 & 30 & 9 & 40 & 10   \\ 
\hline
Total & 27\,400 & 2280  & 21\,350 & 1660 & 14\,130 & 1160 & 8320 & 750 & 6050 & 500 & 5330 & 460   \\ 
Data  & \multicolumn{1}{H}{27\,298} & & \multicolumn{1}{H}{21\,358} & & \multicolumn{1}{H}{14\,157} & & \multicolumn{1}{H}{8361} & & \multicolumn{1}{H}{6066} & & \multicolumn{1}{H}{5385} & \\  
\end{tabular}
}
\label{tab:yields}
\end{table*}

\begin{figure}[!tb]
\centering
\includegraphics[width=\linewidth]{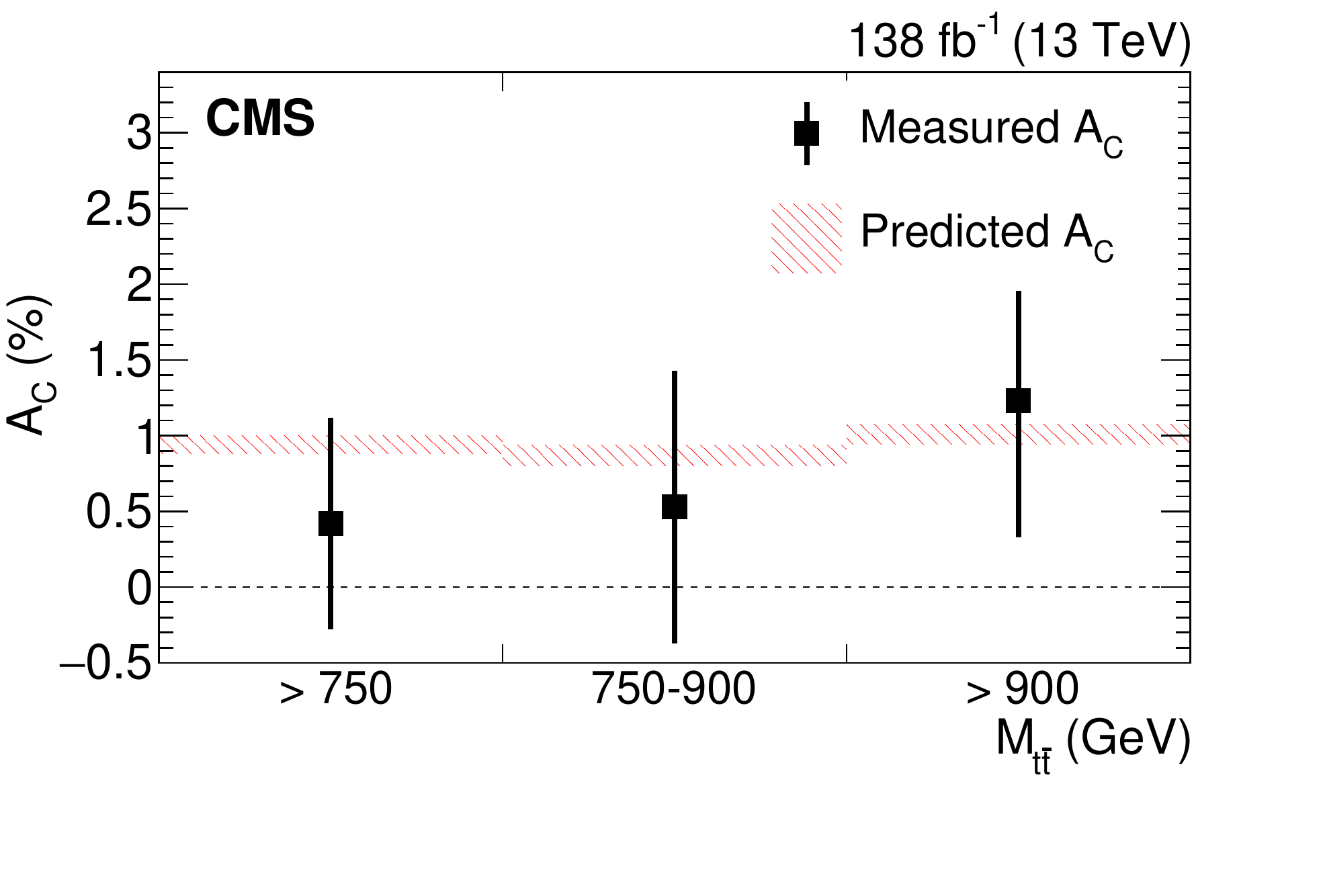}
\caption{Unfolded \AC in the full phase
  space presented in different mass regions after combining
  the \mujets and \ejets channels. The vertical bars represent the
  total uncertainties. The measured values
  are compared to the theoretical prediction, including NNLO
  QCD and NLO EW corrections, from Ref.~\cite{Czakon_2018}. 
  \label{fig:results}}
\end{figure}

Table~\ref{tab:results} and \figrefs{fig:results} show the
measured top quark charge asymmetry after unfolding to parton level in
the full phase space, compared with the theoretical prediction at next-to-next-to-leading order (NNLO) 
QCD and NLO EW corrections from Ref.~\cite{Czakon_2018}. Good
agreement between the data and the MC prediction is
observed. 
   
 \begin{table*}[tbp]
\centering
\topcaption{Measured unfolded charge asymmetry compared to the theoretical prediction, including NNLO
  QCD and NLO EW corrections, from Ref.~\cite{Czakon_2018}. 
   Results are
  shown for events with $\mttbar > 750\GeV$ and for two invariant mass
  ranges, 750--900 and ${>}900 \GeV$. The statistical and
  systematic uncertainties in the data, the MC statistical
  uncertainty, and the total uncertainty in the measured
  values are also shown. All values are in percent.}  
\renewcommand{\arraystretch}{1.3} 
\begin{tabular}{c|cccccc}
$\mttbar (\GeVns)$ & \AC & Stat & Syst  & MC stat & Total & Prediction  \\  \hline
$>$750 & 0.42  &  $\pm {0.44}$ & $_{-0.44}^{+0.33}$ & $\pm \ 0.32$ & $_{-0.69}^{+0.64}$ & $0.94^{ \ +0.05}_{ \ -0.07}$ \\ 
750--900 & 0.53 &  $\pm {0.65}$ & $_{-0.49}^{+0.37}$  & $\pm \ 0.45$ &  $_{-0.93}^{+0.87}$ & $0.87^{ \ +0.06}_{ \ -0.08}$ \\ 
$>$900 &  1.23 &  $\pm {0.58}$  & $_{-0.84}^{+0.43}$  & $\pm \ 0.41$ & $_{-1.10}^{+0.82}$ & $1.01^{ \ +0.06}_{ \ -0.07}$ \\ 
\end{tabular}
\label{tab:results}
\end{table*}

\begin{figure}[htbp]
 \centering
 \includegraphics[width=\linewidth]{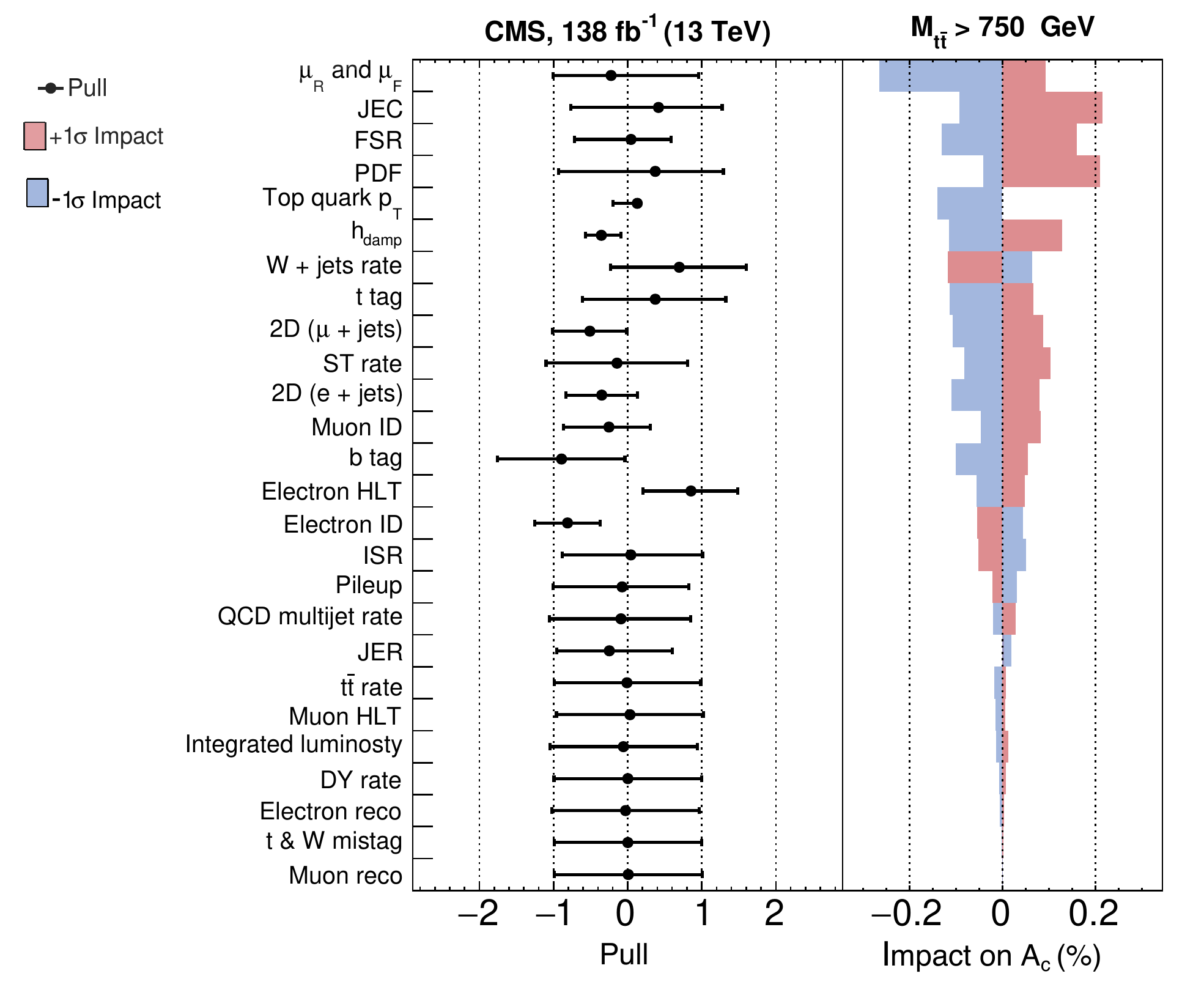}
 \caption{Summary of the dominant systematic uncertainties affecting the $\AC$ measurement. 
 The left column lists the sources of systematic uncertainty, treated as nuisance parameters in the fit, 
 in order of importance. In the middle column (Pull), the black points with their $\pm 1$ standard deviation horizontal bars show for each 
 uncertainty the difference between the observed best fit value and the nominal 
 value, divided by the expected standard deviation. The right column (Impact on $\AC$(\%)), shows 
 the change in the measured $\AC$  if a nuisance is varied  by one standard  deviation ($\sigma$)
 up, in red, and down, in blue. 
   \label{fig:impacts_750_Ac}}
\end{figure}

Figure~\ref{fig:impacts_750_Ac} shows the $\pm1$ standard
deviation ($\sigma$) impacts of the systematic uncertainties
in the \AC measurements for the full signal sample, as well as the effect on the unfolded $\AC$ values for up
and down variations of the systematic uncertainty. 
   
\section{Summary}

A measurement of the charge asymmetry in $\ttbar$ events with
highly boosted top quarks produced in proton-proton collisions at
$\sqrt{s}=13\TeV$ is presented based on 138\fbinv of data collected by
the CMS experiment at the LHC. The selection is optimized for top  
quarks produced with high Lorentz boosts that yield collimated decay
products that are partially or fully merged and can result in
nonisolated leptons and overlapping jets. The measured top quark
charge asymmetry (\AC) is corrected for detector and acceptance
effects using a binned maximum likelihood fit.

This is the first CMS measurement to use 13\TeV data and a binned maximum
likelihood unfolding technique to measure \AC directly at parton level
in the full phase space. In addition, it is the first result that
focuses exclusively on the highly Lorentz-boosted regime, using
dedicated reconstruction techniques for the hadronically and
leptonically decaying top quarks at both the trigger and offline
stages. Since the relative contribution of valence quarks increases at
high momentum transfer, \AC is especially sensitive to beyond the
standard model processes in this highly boosted phase space.
 
The resulting unfolded charge asymmetry for \ttbar events with
invariant masses satisfying $\mttbar > 750\GeV$ is $(0.42_{- 0.69}^{+ 0.64})\%$, where the
uncertainty includes both statistical and systematic components. The
corresponding theoretical prediction at next-to-next-to-leading order
in QCD perturbation theory with next-to-leading-order electroweak
corrections from Ref.~\cite{Czakon_2018} is $(0.94^{+
  0.05}_{-0.07})\%$. Good agreement between the measurement and the
most precise standard model prediction is thus observed. The result
demonstrates that top quark properties can be precisely measured in
the highly boosted topology.

\begin{acknowledgments}
We congratulate our colleagues in the CERN accelerator departments for the excellent performance of the LHC and thank the technical and administrative staffs at CERN and at other CMS institutes for their contributions to the success of the CMS effort. In addition, we gratefully acknowledge the computing centers and personnel of the Worldwide LHC Computing Grid and other centers for delivering so effectively the computing infrastructure essential to our analyses. Finally, we acknowledge the enduring support for the construction and operation of the LHC, the CMS detector, and the supporting computing infrastructure provided by the following funding agencies: BMBWF and FWF (Austria); FNRS and FWO (Belgium); CNPq, CAPES, FAPERJ, FAPERGS, and FAPESP (Brazil); MES and BNSF (Bulgaria); CERN; CAS, MoST, and NSFC (China); MINCIENCIAS (Colombia); MSES and CSF (Croatia); RIF (Cyprus); SENESCYT (Ecuador); MoER, ERC PUT and ERDF (Estonia); Academy of Finland, MEC, and HIP (Finland); CEA and CNRS/IN2P3 (France); BMBF, DFG, and HGF (Germany); GSRI (Greece); NKFIH (Hungary); DAE and DST (India); IPM (Iran); SFI (Ireland); INFN (Italy); MSIP and NRF (Republic of Korea); MES (Latvia); LAS (Lithuania); MOE and UM (Malaysia); BUAP, CINVESTAV, CONACYT, LNS, SEP, and UASLP-FAI (Mexico); MOS (Montenegro); MBIE (New Zealand); PAEC (Pakistan); MES and NSC (Poland); FCT (Portugal); MESTD (Serbia); MCIN/AEI and PCTI (Spain); MOSTR (Sri Lanka); Swiss Funding Agencies (Switzerland); MST (Taipei); MHESI and NSTDA (Thailand); TUBITAK and TENMAK (Turkey); NASU (Ukraine); STFC (United Kingdom); DOE and NSF (USA).
  
\hyphenation{Rachada-pisek} Individuals have received support from the Marie-Curie program and the European Research Council and Horizon 2020 Grant, contract Nos.\ 675440, 724704, 752730, 758316, 765710, 824093, 884104, and COST Action CA16108 (European Union); the Leventis Foundation; the Alfred P.\ Sloan Foundation; the Alexander von Humboldt Foundation; the Belgian Federal Science Policy Office; the Fonds pour la Formation \`a la Recherche dans l'Industrie et dans l'Agriculture (FRIA-Belgium); the Agentschap voor Innovatie door Wetenschap en Technologie (IWT-Belgium); the F.R.S.-FNRS and FWO (Belgium) under the ``Excellence of Science -- EOS" -- be.h project n.\ 30820817; the Beijing Municipal Science \& Technology Commission, No. Z191100007219010; the Ministry of Education, Youth and Sports (MEYS) of the Czech Republic; the Hellenic Foundation for Research and Innovation (HFRI), Project Number 2288 (Greece); the Deutsche Forschungsgemeinschaft (DFG), under Germany's Excellence Strategy -- EXC 2121 ``Quantum Universe" -- 390833306, and under project number 400140256 - GRK2497; the Hungarian Academy of Sciences, the New National Excellence Program - \'UNKP, the NKFIH research grants K 124845, K 124850, K 128713, K 128786, K 129058, K 131991, K 133046, K 138136, K 143460, K 143477, 2020-2.2.1-ED-2021-00181, and TKP2021-NKTA-64 (Hungary); the Council of Science and Industrial Research, India; the Latvian Council of Science; the Ministry of Education and Science, project no. 2022/WK/14, and the National Science Center, contracts Opus 2021/41/B/ST2/01369 and 2021/43/B/ST2/01552 (Poland); the Funda\c{c}\~ao para a Ci\^encia e a Tecnologia, grant CEECIND/01334/2018 (Portugal); the National Priorities Research Program by Qatar National Research Fund; MCIN/AEI/10.13039/501100011033, ERDF ``a way of making Europe", and the Programa Estatal de Fomento de la Investigaci{\'o}n Cient{\'i}fica y T{\'e}cnica de Excelencia Mar\'{\i}a de Maeztu, grant MDM-2017-0765 and Programa Severo Ochoa del Principado de Asturias (Spain); the Chulalongkorn Academic into Its 2nd Century Project Advancement Project, and the National Science, Research and Innovation Fund via the Program Management Unit for Human Resources \& Institutional Development, Research and Innovation, grant B05F650021 (Thailand); the Kavli Foundation; the Nvidia Corporation; the SuperMicro Corporation; the Welch Foundation, contract C-1845; and the Weston Havens Foundation (USA).
\end{acknowledgments}

\bibliography{auto_generated}

\providecommand{\href}[2]{#2}\begingroup\raggedright\begin{thebibliography}{10}%
\makeatletter
\providecommand{\hrefCMSnoop }[0]{\@secondoftwo}%
\makeatother
\providecommand{\doi}{\texttt{doi:}\begingroup \urlstyle{tt}\Url}

\bibitem{Czakon:2013goa}
\hrefCMSnoop {}{M.~Czakon, P.~Fiedler, and A.~Mitov, ``Total top-quark
  pair-production cross section at hadron colliders through
  $\mathcal{O}(\alpha^4_s)$'',} \textit{ Phys. Rev. Lett.} \textbf{ 110} (2013)
  252004,
  \href{http://dx.doi.org/10.1103/PhysRevLett.110.252004}{\doi{10.1103/PhysRevLett.110.252004}},
  \href{http://www.arXiv.org/abs/1303.6254}{\texttt{arXiv:1303.6254}}.

\bibitem{Catani:2019hip}
S.~Catani\hrefCMSnoop {}{ {et~al.}, ``Top-quark pair production at the {LHC}:
  Fully differential {QCD} predictions at {NNLO}'',} \textit{ JHEP} \textbf{
  07} (2019) 100,
  \href{http://dx.doi.org/10.1007/JHEP07(2019)100}{\doi{10.1007/JHEP07(2019)100}},
  \href{http://www.arXiv.org/abs/1906.06535}{\texttt{arXiv:1906.06535}}.

\bibitem{PhysRevLett.120.042001}
\hrefCMSnoop {}{{CDF and D0 Collaborations}, ``Combined forward-backward
  asymmetry measurements in top-antitop quark production at the {Tevatron}'',}
  \textit{ Phys. Rev. Lett.} \textbf{ 120} (2018) 042001,
  \href{http://dx.doi.org/10.1103/PhysRevLett.120.042001}{\doi{10.1103/PhysRevLett.120.042001}},
  \href{http://www.arXiv.org/abs/1709.04894}{\texttt{arXiv:1709.04894}}.

\bibitem{Czakon_2015}
\hrefCMSnoop {}{M.~Czakon, P.~Fiedler, and A.~Mitov, ``Resolving the {Tevatron}
  top quark forward-backward asymmetry puzzle: Fully differential
  next-to-next-to-leading-order calculation'',} \textit{ Phys. Rev. Lett.}
  \textbf{ 115} (2015) 052001,
  \href{http://dx.doi.org/10.1103/physrevlett.115.052001}{\doi{10.1103/physrevlett.115.052001}},
  \href{http://www.arXiv.org/abs/1411.3007}{\texttt{arXiv:1411.3007}}.

\bibitem{Czakon_2018}
M.~Czakon\hrefCMSnoop {}{ {et~al.}, ``Top-quark charge asymmetry at the {LHC}
  and {Tevatron} through {NNLO} {QCD} and {NLO} {EW}'',} \textit{ Phys. Rev. D}
  \textbf{ 98} (2018) 014003,
  \href{http://dx.doi.org/10.1103/physrevd.98.014003}{\doi{10.1103/physrevd.98.014003}},
  \href{http://www.arXiv.org/abs/1711.03945}{\texttt{arXiv:1711.03945}}.

\bibitem{Czakon_2016}
\hrefCMSnoop {}{M.~Czakon, D.~Heymes, and A.~Mitov, ``High-precision
  differential predictions for top-quark pairs at the {LHC}'',} \textit{ Phys.
  Rev. Lett.} \textbf{ 116} (2016) 082003,
  \href{http://dx.doi.org/10.1103/physrevlett.116.082003}{\doi{10.1103/physrevlett.116.082003}},
  \href{http://www.arXiv.org/abs/1511.00549}{\texttt{arXiv:1511.00549}}.

\bibitem{PDF4LHC}
\hrefCMSnoop {}{J.~Rojo {et~al.}, ``The {PDF4LHC} report on {PDFs} and {LHC}
  data: results from {Run I} and preparation for {Run II}'',} \textit{ J. Phys.
  G} \textbf{ 42} (2015) 103103,
  \href{http://dx.doi.org/10.1088/0954-3899/42/10/103103}{\doi{10.1088/0954-3899/42/10/103103}},
  \href{http://www.arXiv.org/abs/1507.00556}{\texttt{arXiv:1507.00556}}.

\bibitem{arXiv:1109.3710}
\hrefCMSnoop {}{J.~A. Aguilar-Saavedra, A.~Juste, and F.~Rubbo, ``Boosting the
  $\ttbar$ charge asymmetry'',} \textit{ Phys. Lett. B} \textbf{ 707} (2012)
  92,
  \href{http://dx.doi.org/10.1016/j.physletb.2011.12.007}{\doi{10.1016/j.physletb.2011.12.007}},
  \href{http://www.arXiv.org/abs/1109.3710}{\texttt{arXiv:1109.3710}}.

\bibitem{arXiv:0709.1652}
\hrefCMSnoop {}{O.~Antunano, J.~H. Kuhn, and G.~Rodrigo, ``Top quarks,
  axigluons and charge asymmetries at hadron colliders'',} \textit{ Phys. Rev.
  D} \textbf{ 77} (2008) 014003,
  \href{http://dx.doi.org/10.1103/PhysRevD.77.014003}{\doi{10.1103/PhysRevD.77.014003}},
  \href{http://www.arXiv.org/abs/0709.1652}{\texttt{arXiv:0709.1652}}.

\bibitem{Frampton:2009rk}
\hrefCMSnoop {}{P.~H. Frampton, J.~Shu, and K.~Wang, ``Axigluon as possible
  explanation for {$\Pp\PAp\rightarrow\ttbar$} forward-backward asymmetry'',}
  \textit{ Phys. Lett. B} \textbf{ 683} (2010) 294,
  \href{http://dx.doi.org/10.1016/j.physletb.2009.12.043}{\doi{10.1016/j.physletb.2009.12.043}},
  \href{http://www.arXiv.org/abs/0911.2955}{\texttt{arXiv:0911.2955}}.

\bibitem{arXiv:9607207}
\hrefCMSnoop {}{J.~L. Rosner, ``Prominent decay modes of a leptophobic
  {$\PZ^\prime$}'',} \textit{ Phys. Lett. B} \textbf{ 387} (1996) 113,
  \href{http://dx.doi.org/10.1016/0370-2693(96)01022-2}{\doi{10.1016/0370-2693(96)01022-2}},
  \href{http://www.arXiv.org/abs/hep-ph/9607207}{\texttt{arXiv:hep-ph/9607207}}.

\bibitem{arXiv:0809.3354}
\hrefCMSnoop {}{P.~Ferrario and G.~Rodrigo, ``Massive color-octet bosons and
  the charge asymmetries of top quarks at hadron colliders'',} \textit{ Phys.
  Rev. D} \textbf{ 78} (2008) 094018,
  \href{http://dx.doi.org/10.1103/physrevd.78.094018}{\doi{10.1103/physrevd.78.094018}},
  \href{http://www.arXiv.org/abs/0809.3354}{\texttt{arXiv:0809.3354}}.

\bibitem{Ferrario:2009bz}
\hrefCMSnoop {}{P.~Ferrario and G.~Rodrigo, ``Constraining heavy colored
  resonances from top-antitop quark events'',} \textit{ Phys. Rev. D} \textbf{
  80} (2009) 051701,
  \href{http://dx.doi.org/10.1103/PhysRevD.80.051701}{\doi{10.1103/PhysRevD.80.051701}},
  \href{http://www.arXiv.org/abs/0906.5541}{\texttt{arXiv:0906.5541}}.

\bibitem{Aguilar-Saavedra:2011bsa}
\hrefCMSnoop {}{J.~A. Aguilar-Saavedra and M.~Perez-Victoria, ``Asymmetries in
  \ttbar production: {LHC} versus {Tevatron}'',} \textit{ Phys. Rev. D}
  \textbf{ 84} (2011) 115013,
  \href{http://dx.doi.org/10.1103/PhysRevD.84.115013}{\doi{10.1103/PhysRevD.84.115013}},
  \href{http://www.arXiv.org/abs/1105.4606}{\texttt{arXiv:1105.4606}}.

\bibitem{Aguilar-Saavedra:2011mor}
\hrefCMSnoop {}{J.~A. Aguilar-Saavedra and M.~Perez-Victoria, ``Simple models
  for the top asymmetry: Constraints and predictions'',} \textit{ JHEP}
  \textbf{ 09} (2011) 097,
  \href{http://dx.doi.org/10.1007/JHEP09(2011)097}{\doi{10.1007/JHEP09(2011)097}},
  \href{http://www.arXiv.org/abs/1107.0841}{\texttt{arXiv:1107.0841}}.

\bibitem{Benbrik:2013lxa}
\hrefCMSnoop {}{R.~Benbrik, C.-H. Chen, and M.~El~Kacimi, ``Colored bosons on
  top {FBA} and angular cross section for \ttbar production'',} \textit{ Phys.
  Lett. B} \textbf{ 725} (2013) 372,
  \href{http://dx.doi.org/10.1016/j.physletb.2013.07.057}{\doi{10.1016/j.physletb.2013.07.057}},
  \href{http://www.arXiv.org/abs/1304.2273}{\texttt{arXiv:1304.2273}}.

\bibitem{Jung:2009pi}
\hrefCMSnoop {}{D.-W. Jung, P.~Ko, J.~S. Lee, and S.-h. Nam, ``Model
  independent analysis of the forward-backward asymmetry of top quark
  production at the {Tevatron}'',} \textit{ Phys. Lett. B} \textbf{ 691} (2010)
  238,
  \href{http://dx.doi.org/10.1016/j.physletb.2010.06.040}{\doi{10.1016/j.physletb.2010.06.040}},
  \href{http://www.arXiv.org/abs/0912.1105}{\texttt{arXiv:0912.1105}}.

\bibitem{ATLAS:2017gkv}
\hrefCMSnoop {}{{ATLAS and CMS Collaborations}, ``Combination of inclusive and
  differential \ttbar charge asymmetry measurements using {ATLAS} and {CMS}
  data at $ \sqrt{s}=7 $ and 8 {TeV}'',} \textit{ JHEP} \textbf{ 04} (2018)
  033,
  \href{http://dx.doi.org/10.1007/JHEP04(2018)033}{\doi{10.1007/JHEP04(2018)033}},
  \href{http://www.arXiv.org/abs/1709.05327}{\texttt{arXiv:1709.05327}}.

\bibitem{PhysRevD.83.034006}
\hrefCMSnoop {}{C.~Zhang and S.~Willenbrock, ``Effective-field-theory approach
  to top-quark production and decay'',} \textit{ Phys. Rev. D} \textbf{ 83}
  (2011) 034006,
  \href{http://dx.doi.org/10.1103/PhysRevD.83.034006}{\doi{10.1103/PhysRevD.83.034006}},
  \href{http://www.arXiv.org/abs/1008.3869}{\texttt{arXiv:1008.3869}}.

\bibitem{SMEFT}
J.~Ellis\hrefCMSnoop {}{ {et~al.}, ``Top, {H}iggs, diboson and electroweak fit
  to the {S}tandard {M}odel effective field theory'',} \textit{ JHEP} \textbf{
  04} (2021) 279,
  \href{http://dx.doi.org/10.1007/JHEP04(2021)279}{\doi{10.1007/JHEP04(2021)279}},
  \href{http://www.arXiv.org/abs/2012.02779}{\texttt{arXiv:2012.02779}}.

\bibitem{ATLAS:2022mlu}
\hrefCMSnoop {}{{ATLAS Collaboration}, ``{Differential $t\bar{t}$ cross-section
  measurements using boosted top quarks in the all-hadronic final state with
  139 fb$^{-1}$ of ATLAS data}'',} 5, 2022.
  \href{http://www.arXiv.org/abs/2205.02817}{\texttt{arXiv:2205.02817}}.

\bibitem{Sirunyan_2019}
\hrefCMSnoop {}{{CMS Collaboration}, ``Search for resonant
  $\mathrm{t}\overline{\mathrm{t}}$ production in proton-proton collisions at
  $\sqrt{s}=13$ {TeV}'',} \textit{ JHEP} \textbf{ 04} (2019) 031,
  \href{http://dx.doi.org/10.1007/JHEP04(2019)031}{\doi{10.1007/JHEP04(2019)031}},
  \href{http://www.arXiv.org/abs/1810.05905}{\texttt{arXiv:1810.05905}}.

\bibitem{hepdata}
\hrefCMSnoop {}{}{HEPD}ata record for this analysis, 2022.
\newblock
  \href{http://dx.doi.org/10.17182/hepdata.127992}{\doi{10.17182/hepdata.127992}}.

\bibitem{CMS:2020cmk}
\hrefCMSnoop {}{{CMS Collaboration}, ``Performance of the {CMS} {L}evel-1
  trigger in proton-proton collisions at $\sqrt{s} = 13$\,{TeV}'',} \textit{
  JINST} \textbf{ 15} (2020) P10017,
  \href{http://dx.doi.org/10.1088/1748-0221/15/10/P10017}{\doi{10.1088/1748-0221/15/10/P10017}},
  \href{http://www.arXiv.org/abs/2006.10165}{\texttt{arXiv:2006.10165}}.

\bibitem{CMS:2016ngn}
\hrefCMSnoop {}{{CMS Collaboration}, ``The {CMS} trigger system'',} \textit{
  JINST} \textbf{ 12} (2017) P01020,
  \href{http://dx.doi.org/10.1088/1748-0221/12/01/P01020}{\doi{10.1088/1748-0221/12/01/P01020}},
\href{http://www.arXiv.org/abs/1609.02366}{\texttt{arXiv:1609.02366}}.
%%CITATION = ARXIV:1609.02366;%%.

\bibitem{CMS:2008xjf}
\hrefCMSnoop {}{{CMS Collaboration}, ``The {CMS} experiment at the {CERN}
  {LHC}'',} \textit{ JINST} \textbf{ 3} (2008) S08004,
  \href{http://dx.doi.org/10.1088/1748-0221/3/08/S08004}{\doi{10.1088/1748-0221/3/08/S08004}}.

\bibitem{CMS:2017yfk}
\hrefCMSnoop {}{{CMS Collaboration}, ``Particle-flow reconstruction and global
  event description with the {CMS} detector'',} \textit{ JINST} \textbf{ 12}
  (2017) P10003,
  \href{http://dx.doi.org/10.1088/1748-0221/12/10/P10003}{\doi{10.1088/1748-0221/12/10/P10003}},
  \href{http://www.arXiv.org/abs/1706.04965}{\texttt{arXiv:1706.04965}}.

\bibitem{CMS:2014pgm}
\hrefCMSnoop {}{{CMS Collaboration}, ``Description and performance of track and
  primary-vertex reconstruction with the {CMS} tracker'',} \textit{ JINST}
  \textbf{ 9} (2014) P10009,
  \href{http://dx.doi.org/10.1088/1748-0221/9/10/P10009}{\doi{10.1088/1748-0221/9/10/P10009}},
  \href{http://www.arXiv.org/abs/1405.6569}{\texttt{arXiv:1405.6569}}.

\bibitem{CMS-TDR-15-02}
\href {http://cds.cern.ch/record/2020886}{{CMS Collaboration}, ``Technical
  proposal for the {Phase-II} upgrade of the {Compact Muon Solenoid}'',} CMS
  Technical Proposal CERN-LHCC-2015-010, CMS-TDR-15-02, 2015.

\bibitem{Cacciari:2008gp}
\hrefCMSnoop {}{M.~Cacciari, G.~P. Salam, and G.~Soyez, ``The anti-\kt jet
  clustering algorithm'',} \textit{ JHEP} \textbf{ 04} (2008) 063,
  \href{http://dx.doi.org/10.1088/1126-6708/2008/04/063}{\doi{10.1088/1126-6708/2008/04/063}},
  \href{http://www.arXiv.org/abs/0802.1189}{\texttt{arXiv:0802.1189}}.

\bibitem{Cacciari:2011ma}
\hrefCMSnoop {}{M.~Cacciari, G.~P. Salam, and G.~Soyez, ``{FastJet} user
  manual'',} \textit{ Eur. Phys. J. C} \textbf{ 72} (2012) 1896,
  \href{http://dx.doi.org/10.1140/epjc/s10052-012-1896-2}{\doi{10.1140/epjc/s10052-012-1896-2}},
\href{http://www.arXiv.org/abs/1111.6097}{\texttt{arXiv:1111.6097}}.
%%CITATION = ARXIV:1111.6097;%%.

\bibitem{Bertolini:2014bba}
\hrefCMSnoop {}{D.~Bertolini, P.~Harris, M.~Low, and N.~Tran, ``Pileup per
  particle identification'',} \textit{ JHEP} \textbf{ 10} (2014) 059,
  \href{http://dx.doi.org/10.1007/JHEP10(2014)059}{\doi{10.1007/JHEP10(2014)059}},
  \href{http://www.arXiv.org/abs/1407.6013}{\texttt{arXiv:1407.6013}}.

\bibitem{CMS:2020ebo}
\hrefCMSnoop {}{{CMS Collaboration}, ``Pileup mitigation at {CMS} in 13 {TeV}
  data'',} \textit{ JINST} \textbf{ 15} (2020) P09018,
  \href{http://dx.doi.org/10.1088/1748-0221/15/09/P09018}{\doi{10.1088/1748-0221/15/09/P09018}},
  \href{http://www.arXiv.org/abs/2003.00503}{\texttt{arXiv:2003.00503}}.

\bibitem{CMS:2019ctu}
\hrefCMSnoop {}{{CMS Collaboration}, ``Performance of missing transverse
  momentum reconstruction in proton-proton collisions at $\sqrt{s} = 13$\,{TeV}
  using the {CMS} detector'',} \textit{ JINST} \textbf{ 14} (2019) P07004,
  \href{http://dx.doi.org/10.1088/1748-0221/14/07/P07004}{\doi{10.1088/1748-0221/14/07/P07004}},
\href{http://www.arXiv.org/abs/1903.06078}{\texttt{arXiv:1903.06078}}.
%%CITATION = ARXIV:1903.06078;%%.

\bibitem{CMS:2011shu}
\hrefCMSnoop {}{{CMS Collaboration}, ``Determination of jet energy calibration
  and transverse momentum resolution in {CMS}'',} \textit{ JINST} \textbf{ 6}
  (2011) P11002,
  \href{http://dx.doi.org/10.1088/1748-0221/6/11/P11002}{\doi{10.1088/1748-0221/6/11/P11002}},
  \href{http://www.arXiv.org/abs/1107.4277}{\texttt{arXiv:1107.4277}}.

\bibitem{Khachatryan_2017}
\hrefCMSnoop {}{{CMS Collaboration}, ``Jet energy scale and resolution in the
  {CMS} experiment in pp collisions at 8 {TeV}'',} \textit{ JINST} \textbf{ 12}
  (2017) P02014,
  \href{http://dx.doi.org/10.1088/1748-0221/12/02/P02014}{\doi{10.1088/1748-0221/12/02/P02014}},
  \href{http://www.arXiv.org/abs/1607.03663}{\texttt{arXiv:1607.03663}}.

\bibitem{CMS:2020poo}
\hrefCMSnoop {}{{CMS Collaboration}, ``Identification of heavy, energetic,
  hadronically decaying particles using machine-learning techniques'',}
  \textit{ JINST} \textbf{ 15} (2020) P06005,
  \href{http://dx.doi.org/10.1088/1748-0221/15/06/P06005}{\doi{10.1088/1748-0221/15/06/P06005}},
  \href{http://www.arXiv.org/abs/2004.08262}{\texttt{arXiv:2004.08262}}.

\bibitem{Larkoski:2014wba}
\hrefCMSnoop {}{A.~J. Larkoski, S.~Marzani, G.~Soyez, and J.~Thaler, ``Soft
  drop'',} \textit{ JHEP} \textbf{ 05} (2014) 146,
  \href{http://dx.doi.org/10.1007/JHEP05(2014)146}{\doi{10.1007/JHEP05(2014)146}},
  \href{http://www.arXiv.org/abs/1402.2657}{\texttt{arXiv:1402.2657}}.

\bibitem{Thaler:2010tr}
\hrefCMSnoop {}{J.~Thaler and K.~Van~Tilburg, ``Identifying boosted objects
  with {N}-subjettiness'',} \textit{ JHEP} \textbf{ 03} (2011) 015,
  \href{http://dx.doi.org/10.1007/JHEP03(2011)015}{\doi{10.1007/JHEP03(2011)015}},
  \href{http://www.arXiv.org/abs/1011.2268}{\texttt{arXiv:1011.2268}}.

\bibitem{CMS:2017wtu}
\hrefCMSnoop {}{{CMS Collaboration}, ``Identification of heavy-flavour jets
  with the {CMS} detector in pp collisions at 13 {TeV}'',} \textit{ JINST}
  \textbf{ 13} (2018) P05011,
  \href{http://dx.doi.org/10.1088/1748-0221/13/05/P05011}{\doi{10.1088/1748-0221/13/05/P05011}},
  \href{http://www.arXiv.org/abs/1712.07158}{\texttt{arXiv:1712.07158}}.

\bibitem{DeepJetJINST2020}
E.~Bols\hrefCMSnoop {}{ {et~al.}, ``Jet flavour classification using
  {DeepJet}'',} \textit{ JINST} \textbf{ 15} (2020) P12012,
  \href{http://dx.doi.org/10.1088/1748-0221/15/12/P12012}{\doi{10.1088/1748-0221/15/12/P12012}},
  \href{http://www.arXiv.org/abs/2008.10519}{\texttt{arXiv:2008.10519}}.

\bibitem{CMS-LUM-17-003}
\hrefCMSnoop {}{{CMS Collaboration}, ``Precision luminosity measurement in
  proton-proton collisions at $\sqrt{s} =$ 13 {TeV} in 2015 and 2016 at
  {CMS}'',} \textit{ Eur. Phys. J. C} \textbf{ 81} (2021) 800,
  \href{http://dx.doi.org/10.1140/epjc/s10052-021-09538-2}{\doi{10.1140/epjc/s10052-021-09538-2}},
  \href{http://www.arXiv.org/abs/2104.01927}{\texttt{arXiv:2104.01927}}.

\bibitem{CMS-PAS-LUM-17-004}
\href {https://cds.cern.ch/record/2621960/}{{CMS Collaboration}, ``{CMS}
  luminosity measurement for the 2017 data-taking period at $\sqrt{s}$ = 13
  {TeV}'',} CMS Physics Analysis Summary CMS-PAS-LUM-17-004, 2018.

\bibitem{CMS-PAS-LUM-18-002}
\href {https://cds.cern.ch/record/2676164/}{{CMS Collaboration}, ``{CMS}
  luminosity measurement for the 2018 data-taking period at $\sqrt{s}$ = 13
  {TeV}'',} CMS Physics Analysis Summary CMS-PAS-LUM-18-002, 2019.

\bibitem{CMS:2021vhb}
\hrefCMSnoop {}{{CMS Collaboration}, ``Measurement of differential \ttbar
  production cross sections in the full kinematic range using lepton+jets
  events from proton-proton collisions at $\sqrt {s}$ = 13 {TeV}'',} \textit{
  Phys. Rev. D} \textbf{ 104} (2021) 092013,
  \href{http://dx.doi.org/10.1103/PhysRevD.104.092013}{\doi{10.1103/PhysRevD.104.092013}},
  \href{http://www.arXiv.org/abs/2108.02803}{\texttt{arXiv:2108.02803}}.

\bibitem{Frixione:2007nw}
\hrefCMSnoop {}{S.~Frixione, P.~Nason, and G.~Ridolfi, ``{A Positive-weight
  next-to-leading-order Monte Carlo for heavy flavour hadroproduction}'',}
  \textit{ JHEP} \textbf{ 09} (2007) 126,
  \href{http://dx.doi.org/10.1088/1126-6708/2007/09/126}{\doi{10.1088/1126-6708/2007/09/126}},
  \href{http://www.arXiv.org/abs/0707.3088}{\texttt{arXiv:0707.3088}}.

\bibitem{POWHEG}
\hrefCMSnoop {}{S.~Alioli, P.~Nason, C.~Oleari, and E.~Re, ``A general
  framework for implementing {NLO} calculations in shower {Monte Carlo}
  programs: the {POWHEG} {BOX}'',} \textit{ JHEP} \textbf{ 06} (2010) 043,
  \href{http://dx.doi.org/10.1007/JHEP06(2010)043}{\doi{10.1007/JHEP06(2010)043}},
  \href{http://www.arXiv.org/abs/1002.2581}{\texttt{arXiv:1002.2581}}.

\bibitem{Alwall:2014hca}
J.~Alwall\hrefCMSnoop {}{ {et~al.}, ``The automated computation of tree-level
  and next-to-leading order differential cross sections, and their matching to
  parton shower simulations'',} \textit{ JHEP} \textbf{ 07} (2014) 079,
  \href{http://dx.doi.org/10.1007/JHEP07(2014)079}{\doi{10.1007/JHEP07(2014)079}},
  \href{http://www.arXiv.org/abs/1405.0301}{\texttt{arXiv:1405.0301}}.

\bibitem{Sjostrand:2014zea}
T.~Sj{\"o}strand\hrefCMSnoop {}{ {et~al.}, ``An introduction to {PYTHIA
  8.2}'',} \textit{ Comput. Phys. Commun.} \textbf{ 191} (2015) 159,
  \href{http://dx.doi.org/10.1016/j.cpc.2015.01.024}{\doi{10.1016/j.cpc.2015.01.024}},
  \href{http://www.arXiv.org/abs/1410.3012}{\texttt{arXiv:1410.3012}}.

\bibitem{Sirunyan_2020}
\hrefCMSnoop {}{{CMS Collaboration}, ``Extraction and validation of a new set
  of {CMS} {\textsc{pythia}}8 tunes from underlying-event measurements'',}
  \textit{ Eur. Phys. J. C} \textbf{ 80} (2020) 4,
  \href{http://dx.doi.org/10.1140/epjc/s10052-019-7499-4}{\doi{10.1140/epjc/s10052-019-7499-4}},
  \href{http://www.arXiv.org/abs/1903.12179}{\texttt{arXiv:1903.12179}}.

\bibitem{Geant4}
\hrefCMSnoop {}{{GEANT4} Collaboration, ``{\GEANTfour} --- a simulation
  toolkit'',} \textit{ Nucl. Instrum. Meth. A} \textbf{ 506} (2003) 250,
  \href{http://dx.doi.org/10.1016/S0168-9002(03)01368-8}{\doi{10.1016/S0168-9002(03)01368-8}}.

\bibitem{NNPDF:2014otw}
\hrefCMSnoop {}{{NNPDF} Collaboration, ``Parton distributions for the {LHC Run
  II}'',} \textit{ JHEP} \textbf{ 04} (2015) 040,
  \href{http://dx.doi.org/10.1007/JHEP04(2015)040}{\doi{10.1007/JHEP04(2015)040}},
  \href{http://www.arXiv.org/abs/1410.8849}{\texttt{arXiv:1410.8849}}.

\bibitem{JHEP07}
\hrefCMSnoop {}{{CMS Collaboration}, ``Search for $\ttbar$ resonances in
  highly-boosted lepton+jets and fully hadronic final stated in proton-proton
  collisions at 13 {TeV}'',} \textit{ JHEP} \textbf{ 07} (2017) 001,
  \href{http://dx.doi.org/10.1007/JHEP07(2017)001}{\doi{10.1007/JHEP07(2017)001}},
  \href{http://www.arXiv.org/abs/1704.03366}{\texttt{arXiv:1704.03366}}.

\bibitem{CMS:2019prd}
\hrefCMSnoop {}{{CMS Collaboration}, ``Measurement of the top quark
  polarization and spin correlations using dilepton final states in
  proton-proton collisions at 13 {TeV}'',} \textit{ Phys. Rev. D} \textbf{ 100}
  (2019) 072002,
  \href{http://dx.doi.org/10.1103/physrevd.100.072002}{\doi{10.1103/physrevd.100.072002}},
  \href{http://www.arXiv.org/abs/1907.03729}{\texttt{arXiv:1907.03729}}.

\bibitem{CZAKON20142930}
\hrefCMSnoop {}{M.~Czakon and A.~Mitov, ``Top++: {A} program for the
  calculation of the top-pair cross-section at hadron colliders'',} \textit{
  Comput. Phys. Commun.} \textbf{ 185} (2014) 2930,
  \href{http://dx.doi.org/10.1016/j.cpc.2014.06.021}{\doi{10.1016/j.cpc.2014.06.021}},
  \href{http://www.arXiv.org/abs/1112.5675}{\texttt{arXiv:1112.5675}}.

\bibitem{CMS:2018mlc}
\hrefCMSnoop {}{{CMS Collaboration}, ``Measurement of the inelastic
  proton-proton cross section at $ \sqrt{s}=13 $ {TeV}'',} \textit{ JHEP}
  \textbf{ 07} (2018) 161,
  \href{http://dx.doi.org/10.1007/JHEP07(2018)161}{\doi{10.1007/JHEP07(2018)161}},
  \href{http://www.arXiv.org/abs/1802.02613}{\texttt{arXiv:1802.02613}}.

\bibitem{CMS-DP-2018-058}
\href {https://cds.cern.ch/record/2646773}{{CMS Collaboration}, ``Performance
  of the {DeepJet} b tagging algorithm using 41.9\fbinv of data from
  proton-proton collisions at 13 {TeV} with the {Phase 1 CMS} detector'',} CMS
  Detector Performance Report CMS-DP-2018-058, 2018.

\bibitem{NNPDF:2017mvq}
\hrefCMSnoop {}{{NNPDF} Collaboration, ``{Parton distributions from
  high-precision collider data}'',} \textit{ Eur. Phys. J. C} \textbf{ 77}
  (2017) 663,
  \href{http://dx.doi.org/10.1140/epjc/s10052-017-5199-5}{\doi{10.1140/epjc/s10052-017-5199-5}},
  \href{http://www.arXiv.org/abs/1706.00428}{\texttt{arXiv:1706.00428}}.

\bibitem{Combine}
\href {https://cdsweb.cern.ch/record/1379837}{{ATLAS, CMS, and the LHC Higgs
  Combination Group}, ``{Procedure for the LHC Higgs boson search combination
  in Summer 2011}'',} {CMS/ATLAS} note CMS-NOTE-2011-005, ATL-PHYS-PUB-2011-11,
  2011.

\bibitem{BARLOW1993219}
\hrefCMSnoop {}{R.~J. Barlow and C.~Beeston, ``Fitting using finite {Monte
  Carlo} samples'',} \textit{ Comput. Phys. Commun.} \textbf{ 77} (1993) 219,
  \href{http://dx.doi.org/10.1016/0010-4655(93)90005-W}{\doi{10.1016/0010-4655(93)90005-W}}.

\bibitem{Combine-github}
\href {https://cms-analysis.github.io/HiggsAnalysis-CombinedLimit/}{{CMS
  Collaboration}, ``Combine: Software tools used for statistical analysis'',}
  2020.
\newblock \url {https://cms-analysis.github.io/HiggsAnalysis-CombinedLimit/}.

\end{thebibliography}\endgroup
\cleardoublepage \appendix\section{The CMS Collaboration \label{app:collab}}\begin{sloppypar}\hyphenpenalty=5000\widowpenalty=500\clubpenalty=5000
\cmsinstitute{Yerevan Physics Institute, Yerevan, Armenia}
{\tolerance=6000
A.~Tumasyan\cmsAuthorMark{1}\cmsorcid{0009-0000-0684-6742}
\par}
\cmsinstitute{Institut f\"{u}r Hochenergiephysik, Vienna, Austria}
{\tolerance=6000
W.~Adam\cmsorcid{0000-0001-9099-4341}, J.W.~Andrejkovic, T.~Bergauer\cmsorcid{0000-0002-5786-0293}, S.~Chatterjee\cmsorcid{0000-0003-2660-0349}, K.~Damanakis\cmsorcid{0000-0001-5389-2872}, M.~Dragicevic\cmsorcid{0000-0003-1967-6783}, A.~Escalante~Del~Valle\cmsorcid{0000-0002-9702-6359}, P.S.~Hussain\cmsorcid{0000-0002-4825-5278}, M.~Jeitler\cmsAuthorMark{2}\cmsorcid{0000-0002-5141-9560}, N.~Krammer\cmsorcid{0000-0002-0548-0985}, L.~Lechner\cmsorcid{0000-0002-3065-1141}, D.~Liko\cmsorcid{0000-0002-3380-473X}, I.~Mikulec\cmsorcid{0000-0003-0385-2746}, P.~Paulitsch, F.M.~Pitters, J.~Schieck\cmsAuthorMark{2}\cmsorcid{0000-0002-1058-8093}, R.~Sch\"{o}fbeck\cmsorcid{0000-0002-2332-8784}, D.~Schwarz\cmsorcid{0000-0002-3821-7331}, M.~Sonawane\cmsorcid{0000-0003-0510-7010}, S.~Templ\cmsorcid{0000-0003-3137-5692}, W.~Waltenberger\cmsorcid{0000-0002-6215-7228}, C.-E.~Wulz\cmsAuthorMark{2}\cmsorcid{0000-0001-9226-5812}
% ...
\par}
\vskip\cmsinstskip
\dag:~Deceased\\
$^{1}$Also at Yerevan State University, Yerevan, Armenia\\
$^{2}$Also at TU Wien, Vienna, Austria\\
% ...
\end{sloppypar}
%%% END EDITABLE REGION %%%
% skeleton_end
\end{document}